\documentclass[a4paper,12pt]{article}

\usepackage{graphicx}
\usepackage{amsfonts}
\usepackage{caption}
\usepackage{amsmath} 
\usepackage{amsthm}
\usepackage{mathtools}
\usepackage{dsfont}
\usepackage{physics}
\usepackage{bm}
\usepackage{enumitem}
\usepackage[svgnames]{xcolor}
\usepackage{pdflscape}
\usepackage{float}
\usepackage{url}
\usepackage{hyperref}
\usepackage{newfloat}
\usepackage{afterpage} 

\DeclareFloatingEnvironment[
    name=Table,           
    placement=htbp,       
    fileext=lot,         
    listname={List of Tables}  
]{Table}

\DeclareFloatingEnvironment[
    name=Code,           
    placement=htbp,       
    fileext=loc,         
    listname={List of Included Code}  
]{code}

\usepackage{enumitem}
\usepackage[margin=1in]{geometry}

\usepackage[onehalfspacing]{setspace}

\allowdisplaybreaks 
\usepackage{soul} 
\usepackage{booktabs} 
\usepackage{pdflscape} 

\usepackage[
    backend=biber,
    style=numeric
]{biblatex}
\newcommand{\pr}{\text{Pr}}

\newcommand{\iid}{\overset{iid}{\sim}}

\newcommand{\uniform}{\text{Uniform}}
\newcommand{\bernoulli}{\text{Bernoulli}}

\newcommand{\ind}{\mathds{1}}

\newcommand{\textref}{\text{ref}}
\newcommand{\llod}{{(\ell)}}

\newcommand{\variant}{{(v)}}
\newcommand{\varianti}{{(v_i)}}
\newcommand{\dvar}{{(d)}}
\newcommand{\ovar}{{(o)}}

\title{Bayesian calendar-time survival analysis with epidemic curve priors and variant-specific infection hazards}

\author{\normalsize Angela M Dahl$^{1,2}$ and Elizabeth R Brown$^{1,2,3}$ \\
\footnotesize $^1$Department of Biostatistics, University of Washington, Seattle, Washington, U.S.A. \\
\footnotesize $^2$Vaccine and Infectious Disease Division, Fred Hutch Cancer Center, Seattle, Washington, U.S.A. \\
\footnotesize $^3$Public Health Sciences Division, Fred Hutch Cancer Center, Seattle, Washington, U.S.A.}

\date{}

\begin{document}

\maketitle

\begin{abstract}
\normalsize
In this paper, we develop a Bayesian calendar-time survival model motivated by infectious disease prevention studies occurring during an epidemic, when the risk of infection can change rapidly as the epidemic curve shifts. For studies in which a biomarker is the predictor of interest, we include the option to estimate a threshold of protection for the biomarker. If the intervention is hypothesized to have different associations with several circulating viral variants, or if the infectiousness of the dominant variant(s) changes over the course of the study, we treat infection from different variants as competing risks. We also introduce a novel method for incorporating existing epidemic curve estimates into an informative prior for the baseline hazard function, enabling estimation of the intervention’s association with infection risk during periods of calendar time with minimal follow-up in one or more comparator groups. We demonstrate the strengths of this method via simulations, and we apply it to data from an observational COVID-19 vaccine study.
\end{abstract}

\begin{center}

\vspace{0.2in}

\small \textbf{Corresponding author} 

\normalsize Angela M Dahl \\
\url{adahl@fredhutch.org} \\
(507) 990-5709

\end{center}

\newpage
\section{Introduction} \label{survival:sec:intro}

In studies of infectious disease prevention, assessing the association between an intervention and the risk of infection is frequently complicated by temporal (often seasonal) changes in the risk of exposure to infection. This is relevant, for example, in vaccine trials for diseases such as influenza, coronavirus 2019 (COVID-19), or respiratory syncytial virus (RSV). For prospective studies with time-to-event outcomes, the estimand of interest is often a hazard ratio obtained from a Cox proportional hazards model, which naturally incorporates right censoring, allows for stratification by site, handles potential confounders via covariate adjustment, and allows for fluctuations in the baseline risk of infection over the course of the study.\supercite{cox_1972, cox_1975, aalen2015does-f92} If a study is randomized and participants are enrolled quickly, then the risk of infection is typically modeled using a study time scale (for example, time since vaccination), as randomization ensures that the average baseline risk of infection is approximately equal between intervention groups at a given time on study, provided there is minimal unobserved heterogeneity in the risk of infection among the study population.\supercite{fintzi_follman_2021} However, in nonrandomized trials, or when participants are enrolled over a period of time rather than all at once, it is standard to model infection risk in calendar time, rather than study time, to account for significant changes in the risk of infection as the background epidemic changes over calendar time, which may otherwise lead to biased comparisons. Fintzi and Follman\supercite{fintzi_follman_2021} illustrate the importance of using calendar time in infectious disease studies with staggered enrollment in order to maintain the proportional hazards assumption; otherwise, if time on study is used as the time scale, then the intervention effect would be biased from the calendar time differences in the baseline hazard between intervention groups. In the context of COVID-19, several authors have highlighted the presence of calendar time bias in observational COVID-19 vaccine studies,\supercite{ioannidis_2022, fung_2024} and many studies of COVID-19 vaccine effectiveness and durability have used calendar time in their models for infection risk, either as the time scale for a Cox proportional hazards model \supercite{covid_vaccine_effectiveness_denmark, covid_vaccine_netherlands} or as a covariate in a logistic or Poisson regression model.\supercite{covid_vaccine_durability, covid_vaccine_scotland_brazil} In a simulated observational COVID-19 vaccine study closely resembling the 2021 pandemic in Denmark, Lund et al\supercite{covid_calendar_time_cox} found that Cox regression with a calendar time scale yielded the least biased results. 

While the standard Cox proportional hazards model is easily applied to infectious disease studies by using a calendar time scale, issues can still arise from its treatment of the baseline hazard function as a nuisance parameter.\supercite{cox_1975} On the one hand, the partial likelihood approach implicitly allows for a completely flexible baseline hazard function, which can handle the rapidly changing infection risk during an epidemic. However, with a flexible baseline hazard function comes a tradeoff: if there are any periods of calendar time with minimal study follow-up in one or more comparator groups, as is common in trials with staggered enrollment, then the model has no information to distinguish the intervention effect from changes in the baseline hazard during these time periods, and the estimated intervention effect may be biased. For example, consider an observational vaccine study that begins enrolling participants in one dose group before another dose group. If the epidemic is at a high point during the beginning of the study but declines by the time the second dose group begins enrolling, then the model may attribute the high rate of infections at the beginning of the study to low protection against infection in the first dose group rather than an especially high baseline hazard during that time. To solve this problem, we propose a Bayesian approach to a calendar-time survival model, enabling us to employ an informative baseline hazard prior to provide the model with information about infection risk during time periods that lack sufficient study data. We propose using publicly available data about the epidemic curve to create this informative prior.

The method we develop in this paper addresses several issues that can arise in trials for infectious diseases that occur in seasonal or epidemic patterns. First, as discussed above, the hazard of infection should be modeled in calendar time, particularly if the trial is nonrandomized or has a long enrollment period. Second, the baseline hazard function should have the flexibility to change rapidly to reflect the behavior of epidemic curves. Third, we could ideally incorporate prior information into the model to inform infection risk during calendar time periods lacking sufficient follow-up. Fourth, for analyses in which a biomarker is the predictor of interest, we would like the option to estimate a threshold of protection. Finally, the model should account for multiple circulating variants of the pathogen if the intervention is hypothesized to be associated with different levels of protection from each variant. 

To address the need for a flexible baseline hazard function, our method uses a piecewise constant baseline hazard function, as typical parametric forms for the baseline hazard such as exponential or Weibull cannot capture the rapidly changing infection risk during an epidemic. However, as noted above, using a flexible baseline hazard function in a calendar time model can lead to insufficient information to identify the intervention effect from the baseline hazard during time periods with little follow-up. We address this problem with an informative baseline hazard prior.

Our method utilizes existing epidemic curve data to derive an informative prior for the baseline hazard function. Our motivation for doing so comes from the assumption that infection risk depends on calendar time only through the changing risk of \textit{exposure} to the pathogen. Following this, we expect the epidemic curve to reflect study participants' risk of exposure on a relative scale; if the rate of infections in the community increases tenfold over the course of a month, then we expect study participants' risk of exposure to infection to also increase approximately tenfold. Once exposed, we assume that participants' risk of becoming infected no longer depends on the current size of the epidemic, but rather on characteristics of their own immune systems and the infectiousness of the variant to which they were exposed. These assumptions highlight the framework in which we think of the infection process: similar to the approach taken by Halloran et al,\supercite{halloran_VE_exposure} we conceive of any infection as resulting from 1) exposure to infection, and 2) infection once exposed (i.e., acquisition). Tsiatis and Davidian\supercite{tsiatis_2021} used a similar framework for conceptualizing vaccine efficacy during the COVID-19 pandemic.

Assuming the probability of infection given no exposure is zero and the probability of exposure is independent of whether someone has been infected previously or not, we can (informally) decompose the hazard of infection as
    \begin{align*}
        &\pr(\text{infection at time $t$ $|$ no infection before time $t$}) \\
        = \; &\pr(\text{infection at time $t$ $|$ exposure at time $t$, no infection before time $t$}) \\
        &\times \pr(\text{exposure at time $t$}). 
    \end{align*}
In our method, we model the probability of exposure as a piecewise constant function that depends on calendar time, and we model the hazard of acquisition using a standard proportional hazards model\supercite{cox_1972} with time-varying covariates. The epidemic curve will then serve as an informative prior for the probability of exposure.

We cannot directly use the epidemic curve itself as a prior, as the rate of infections per person in the epidemic curve data is unlikely to be on the same scale as in our study. Instead, we use the \textit{shape} of the epidemic curve to derive the \textit{shape} of the prior for the baseline hazard function while allowing the \textit{scale} of the baseline hazard function to be driven by the rate of infections in our study. We accomplish this by choosing some reference point in time to which we anchor the scale of the baseline hazard, and then we put informative priors only on the \textit{relative} baseline hazard at other calendar times compared to the reference time. 

Many studies of infectious diseases are interested in the association between a biomarker, such as vaccine-elicited antibodies, and the risk of infection; however, many biomarker assays have a lower limit of detection (LLOD) below which we cannot obtain a reliable measurement.\supercite{llod} Additionally, we are often interested in estimating a threshold of protection, which can guide dosage recommendations for vaccines or monoclonal antibodies.\supercite{thresholds_of_protection} If the predictor of interest is a biomarker with a LLOD, we include a binary indicator in the hazard function denoting whether the biomarker is above or below the LLOD. This allows estimation of the continuous association between biomarker levels and infection risk only among biomarkers above the LLOD. If the goal is to estimate a threshold of protection, we replace this binary LLOD indicator with an indicator denoting whether the biomarker is above the threshold, and we estimate this threshold as an additional parameter in the model. In both approaches, the model assumes a constant risk of infection below the LLOD or threshold of protection at a given calendar time, conditional on any time-varying covariates.

When necessary, we treat multiple variants of the pathogen as competing risks.\supercite{prentice_competing_risks} This is relevant in studies where multiple variants are circulating that change in prevalence over time and have different levels of infectiousness, causing the risk of infection to change over calendar time given the same level of exposure; as Tsiatis and Davidian\supercite{tsiatis_2021} note, the rapidly mutating severe acute respiratory syndrome coronavirus 2 (SARS-CoV-2) variants provide a well-known recent example of this.\supercite{alpha_vs_delta_transmission, delta_vs_omicron_transmission} In addition, this allows us to estimate the association between the predictor of interest and infection risk separately for each variant, which is relevant for many viruses that mutate quickly, such as SARS-CoV-2,\supercite{covid_vaccine_variant_escape, covid_vaccine_antibodies_variants} influenza,\supercite{flu_vaccine_variants} and RSV\supercite{rsv_genomic_diversity}; vaccines for such viruses are typically developed to elicit antibodies against a specific variant, and they may be less effective against other variants that are less similar to the targeted variant. This approach for estimating variant-specific vaccine protection via competing risks is a form of sieve analysis.\supercite{gilbert_sieve_analysis, gilbert_sieve_analysis_examples}


The rest of this paper is organized as follows. In Section \ref{survival:sec:methods}, we derive the likelihood for our model, and we describe our method for incorporating epidemic curve data into an informative prior for the baseline hazard function. In Section \ref{survival:sec:simulations}, we demonstrate our method through simulations, including a demonstration of the bias correction that results from using an informative baseline hazard prior in trials with staggered enrollment. In Section \ref{survival:sec:momivax_analysis}, we apply our method to the MOMI-Vax study, an observational study of SARS-CoV-2 vaccination in pregnant women. Finally, in Section \ref{survival:sec:discussion}, we summarize the benefits and drawbacks of our method, and we discuss the results of our MOMI-Vax analysis.

\section{Methods} \label{survival:sec:methods}

In this section, we derive the likelihood for our model, which incorporates the predictor of interest and any baseline or time-varying covariates via a proportional hazards model. To enable estimation of variant-specific associations between the predictor of interest and the risk of infection, and to allow for changes in the pathogen's infectiousness over calendar time due to new variants, we formulate our survival model as a competing risks model by defining cause-specific hazards for each variant.\supercite{prentice_competing_risks} As introduced in Section \ref{survival:sec:intro}, we express the variant-specific hazard of infection as the product of the hazard of infection \textit{given exposure} to that variant, which we call the hazard of acquisition, and the probability of exposure to that variant. We then derive the forms for the hazard of acquisition, which is a standard proportional hazards model, and the probability of exposure, which is a piecewise constant function that depends on calendar time. Finally, we describe a new method for creating an informative prior for the baseline hazard function by using epidemic curve data to inform the baseline probability of exposure as it changes over calendar time.

Let $V$ be the total number of variants of interest that can infect participants in the study (for many studies, $V=1$), and let $\bm{\theta}$ denote the set of model parameters. For participant $i$, let $t_{i0}$ denote the calendar time at which their study follow-up begins. At calendar time $t$, let $h^\variant_{A, i}(t | \bm{\theta})$ denote the hazard of acquisition for variant $v$, and let $p^\variant_{E,i}(t | \bm{\theta})$ denote the probability of exposure to variant $v$. Then the cause-specific hazard of infection from variant $v$ can be written as $h^\variant_i(t | \bm{\theta}) = h^\variant_{T,i}(t | \bm{\theta}) \times p^\variant_{E,i}(t | \bm{\theta}),$ and the overall hazard of infection (from any variant) is $h_i(t | \bm{\theta}) = \sum_{v=1}^V h^\variant_i(t | \bm{\theta})$. In the subsections that follow, we describe the models for the hazard of transmission and the probability of exposure, which combine to get the hazard of infection.

\subsection{Hazard of acquisition} \label{survival:sec:acquisition_hazard}

We model the hazard of acquisition (that is, the hazard of infection given exposure) as a function of calendar time $t$ depending on each participant's (possibly time-varying) predictor of interest $X_i(t)$ and susceptibility covariates $\bm{Z}_{A,i}(t)$. We assume that, conditional on the predictor of interest and the susceptibility covariates, the hazard of acquisition is constant over time for a given variant. We employ a proportional hazards model, assuming the covariates have the same association with all variants but allowing the predictor of interest to have variant-specific associations with the hazard of acquisition. We include a parameter $\alpha^\variant$ to capture the relative infectiousness of variant $v$ compared to some reference variant $v=1$, with $\alpha^{(1)} = 0$. Then the variant-specific hazard of acquisition for variant $v$ at calendar time $t$ is
    \begin{equation} \label{survival:eq:h_i_e}
        h^\variant_{A, i}(u | \bm{\theta}) = h_{A,0} \exp \Big\{\alpha^\variant + \gamma^\variant X_i(t) + \bm{Z}_{A,i}(t) \bm{\beta}_{A} \Big\},
    \end{equation}
where $h_{A,0}$ is the constant baseline hazard of acquisition.

If the predictor of interest is a biomarker, we can optionally incorporate a LLOD or threshold of protection $X_T$ into the variant-specific hazard of acquisition as follows, assuming a constant risk of infection below $X_T$ at a given calendar time (conditional on any covariates):
    \begin{equation} \label{survival:eq:h_i_e_threshold}
    \begin{aligned}
        h^\variant_{A, i}(u | \bm{\theta}) = h_{A,0} \exp \Big\{\alpha^\variant &+ \gamma^\variant X_i(t) \ind\{X_i(t) > X_T\} + \gamma^\variant_T \ind\{X_i(t) > X_T\} + \bm{Z}_{A,i}(t) \bm{\beta}_{A} \Big\},
    \end{aligned}
    \end{equation}
The threshold $X_T$ may be treated as a parameter itself, allowing estimation of a threshold of protection.

\subsection{Probability of exposure} \label{survival:sec:pr_exposure}

We model the probability of exposure as a function of calendar time $t$ depending on some (possibly time-varying) exposure covariates $\bm{Z}_{E,i}(t)$. For a participant at site $s_i$, the probability of exposure to variant $v$ depends on their instantaneous rate of contacts $c_i(t)$ and the probability a contact at that site is infectious with that variant $p^\variant_{c,s_i}(t)$.\supercite{halloran_VE_exposure} We assume that, conditional on the exposure covariates, participants mix randomly with their community at a constant rate over time, so the contact rate is $c_i(t) = c_0 \exp \big\{ \bm{Z}_{E,i}(t) \bm{\beta}_E \big\}$, where $c_0$ is the baseline contact rate. We assume that the probability a contact is infectious with a particular variant depends on calendar time due to the changing background rate of infections in the community as well as the changing prevalence of that variant in the community. We make use of existing variant surveillance data and treat the relative proportions of each variant in circulation as known piecewise constant functions, denoted $\pi_{s_i}^{(v)}(t)$ for variant $v$, with $\sum_{v=1}^V \pi_{s_i}^\variant(t) = 1$ for all $t$. Then the probability a contact at site $s_i$ is infectious with variant $v$ is $p^\variant_{c,s_i}(t) = \pi^\variant_{s_i}(t) \; p_{c,0,s_i}(t)$. We approximate the probability that a contact at site $s_i$ is infectious with \textit{any} variant using a piecewise constant function, denoted $p_{c,0,s_i}(t)$, over $K$ equally-sized intervals of calendar time $(t_0, t_1], \dots, (t_{K-1}, t_K]$. Then the probability of exposure to infection from variant $v$ is the product of their contact rate and the probability a contact is infectious with variant $v$:
    \begin{equation} \label{survival:eq:p_e}
        p^\variant_{E,i}(t | \bm{\theta}) = c_i(t) \; p^\variant_{c, s_i}(t) = c_0 \; \pi_{s_i}^\variant(t) \; p_{c,0,s_i}(t) \exp \big\{ \bm{Z}_{E,i}(t) \bm{\beta}_E \big\}.
    \end{equation}

\subsection{Hazard of infection} \label{survival:sec:overall_hazard}

Taking the product of our expressions for the hazard of acquisition $h^\variant_{A,i}(t | \bm{\theta})$ and the probability of exposure $p^\variant_{E,i}(t | \bm{\theta})$ for each variant, the variant-specific hazard of infection for variant $v$ is
    \begin{equation}
    \begin{aligned}
        h^\variant_i(t | \bm{\theta}) &= h_{A,0} \exp \Big\{ \alpha^\variant + \gamma^\variant X_i(t) + \bm{Z}_{A,i}(t) \bm{\beta}_{A} \Big\} \\
        &\hphantom{=} \times c_0 \; \pi_{s_i}^\variant(t) \; p^{c,0,s_i}(t) \exp \big\{ \bm{Z}_{E,i}(t) \bm{\beta}_E \big\}.
    \end{aligned}
    \end{equation}
If the model includes a LLOD or threshold for the predictor of interest, then the variant-specific hazard of infection is
    \begin{equation}
    \begin{aligned}
        h^\variant_i(t | \bm{\theta}) &= h_{A,0} \exp \Big\{ \alpha^\variant + \gamma^\variant X_i(t) \ind\{X_i(t) > X_T\} + \gamma^\variant_T \ind\{X_i(t) > X_T\} + \bm{Z}_{A,i}(t) \bm{\beta}_{A} \Big\} \\
        &\hphantom{=} \times c_0 \; \pi_{s_i}^\variant(t) \; p^{c,0,s_i}(t) \exp \big\{ \bm{Z}_{E,i}(t) \bm{\beta}_E \big\},
    \end{aligned}
    \end{equation}
Because we only observe infections, but not exposures and acquisition events separately, the parameters $\bm{\beta}_{A}$, $\bm{\beta}_E$, $h_{A,0}$, $c_0$, and $p_{c,0,s_i}(t)$ are not identifiable on their own. Instead, we combine them into a single vector of coefficients $\bm{\beta}$ and a single piecewise constant baseline hazard term $h^\variant_{0,s_i}(t) = h_{A,0} \; c_0 \; p_{c,0,s_i}(t)$. Keep in mind that the baseline hazard depends on calendar time only through $p_{c,0,s_i}(t)$, the probability a contact at site $s_i$ is infectious. We also cannot typically rule out unmeasured confounding in the relationship between the predictor of interest and the risk of infection, unless the study is randomized; therefore, $\gamma^\variant$ and $\gamma^\variant_T$ actually estimate the association between the predictor of interest and the threshold of protection (if applicable) and the hazard of infection overall, not on the hazard of acquisition. Then the final form of the variant-specific hazard of infection is
    \begin{equation} \label{survival:eq:variant_specific_hazard}
        h^\variant_i(t | \bm{\theta}) = \pi_{s_i}^\variant(t) \; h^\variant_{0, s_i}(t) \exp \Big\{ \alpha^\variant + \gamma^\variant X_i(t) + \bm{Z}_{i}(t) \bm{\beta} \Big\}.
    \end{equation}
If the model includes a LLOD or threshold of protection for the predictor of interest, then the final form of the variant-specific hazard of infection is
    \begin{equation} \label{survival:eq:variant_specific_hazard_threshold}
        h^\variant_i(t | \bm{\theta}) = \pi_{s_i}^\variant(t) \; h^\variant_{0, s_i}(t) \exp \Big\{ \alpha^\variant + \gamma^\variant X_i(t) \ind\{X_i(t) > X_T\} + \gamma^\variant_T \ind\{X_i(t) > X_T\} + \bm{Z}_{i}(t) \bm{\beta} \Big\}.
    \end{equation}

\subsection{Likelihood} \label{survival:sec:llk}

Following Prentice et al,\supercite{prentice_competing_risks} we express the likelihood in terms of the variant-specific hazard functions. For uncensored or right censored participants, let $T_i$ denote their survival time and $\Delta_i$ indicate right censoring. For interval censored participants, let $(L_i, R_i)$ denote their censoring interval. In Section \ref{survival:sec:overall_hazard}, we defined the variant-specific hazard function $h^\variant_{i}(t | \bm{\theta})$. Now, assuming it is only possible to be infected with one variant at a time, the overall hazard function is defined as the sum of the variant-specific hazard functions
\begin{equation}
    h_i(t | \bm{\theta}) = \sum_{v=1}^V h^\variant_{i}(t | \bm{\theta}),
\end{equation}
and the overall survivor function is defined as
\begin{equation}
    S_i(t | \bm{\theta}) = \exp \Big\{ - \int_{t_{i0}}^t h_i(u |\bm{\theta}) du \Big\} = \exp \Big\{ - \int_{t_{i0}}^t \sum_{v=1}^V h^\variant_i(u |\bm{\theta}) du \Big\}.
\end{equation}

\subsubsection{Known infecting variants} \label{survival:sec:llk_known_variants}

For uncensored or right censored participants with infecting variant $v_{i}$ ($v_i$ is arbitrary for right censored participants), the likelihood contribution for their survival time $T_i$, right censoring indicator $\Delta_i$, and infecting variant $v_i$ 
\begin{equation}
    L(T_i, \Delta_i | \bm{\theta}, v_i ) = \Big[ h^\varianti_{i}(T_i | \bm{\theta}) \Big]^{\Delta_i} \times S_i(T_i | \bm{\theta}).
\end{equation}
For interval censored participants, the likelihood contribution for their censoring interval $(L_i, R_i)$ and infecting variant $v_i$ is
\begin{equation}
\begin{aligned}
    L(L_i, R_i | \bm{\theta}, v_i ) &= \int_{L_i}^{R_i} \Big[ h^\varianti_{i}(t | \bm{\theta}) \times S_i(t | \bm{\theta}) \Big] dt.
\end{aligned}
\end{equation}


\subsubsection{Unknown infecting variants} \label{survival:sec:llk_unknown_variants}

If the infecting variants are unknown, the likelihood for each observed infection uses the overall hazard function instead of the variant-specific hazard function. For uncensored or right censored participants, the likelihood contribution for their survival time $T_i$ and right censoring indicator $\Delta_i$ is
\begin{equation}
    L(T_i, \Delta_i | \bm{\theta} ) = \Big[ h_{i}(T_i | \bm{\theta}) \Big]^{\Delta_i} \times S_i(T_i | \bm{\theta}).
\end{equation}
For interval censored participants, the likelihood contribution for their censoring interval $(L_i, R_i)$ is
\begin{equation}
    L(L_i, R_i | \bm{\theta}) = \int_{L_i}^{R_i} \Big[ h_{i}(t | \bm{\theta}) \times S_i(t | \bm{\theta}) \Big] dt = S_i(L_i | \bm{\theta}) - S_i(R_i | \bm{\theta}).
\end{equation}

\subsection{Informative baseline hazard prior based on an epidemic curve} \label{survival:sec:bl_hazard_prior}

We now derive an informative prior for the baseline hazard function that utilizes existing epidemic curve estimates. An informative prior adds information to the model about infection risk during calendar times with little follow-up, improving identifiability of $\gamma$ and $\beta$. Our motivation for using epidemic curve data comes from our formulation of the hazard function, in which we isolated the probability a contact is infectious as the only parameter in the hazard function that depends on calendar time. The probability a contact is infectious constitutes a portion of the baseline hazard function that comes from the probability of exposure. As long as the epidemic curve well represents the background rate of infections in the community with which study participants are mixing, it should be approximately proportional to the probability a study participant's contact is infectious over calendar time, though it will not necessarily be on the same scale. Leveraging this assumption, we use the epidemic curve as an informative prior for the \textit{shape} of the baseline hazard function while using a weakly informative prior for its \textit{scale}. We assume throughout that the epidemic curve corresponds to the overall number of infections and is not separated by variant.

Recall that the baseline hazard of infection at study site $s_i$ is a piecewise constant function over $K$ equally-sized intervals of calendar time $(t_0, t_1], \dots, (t_{K-1}, t_K]$, and it is made up of three terms: the baseline hazard of acquisition $h_{A,0}$, the baseline contact rate $c_0$, and the probability a contact is infectious with $p_{c,0,s_i}(t)$. Because these terms are not identifiable on their own, we combine them into a single baseline hazard term $h_{0, s_i}(t)$. In order to more readily put a prior on the shape of the baseline hazard function, we now reparameterize it as $h_{0, s_i}(t) = h_{\textref, s_i} \times r_{s_i}(t)$, where $h_{\textref, s_i}$ is the baseline hazard during some reference interval of time and $r_{s_i}(t)$ is the relative baseline hazard at calendar time $t$ compared to the reference interval. The relative baseline hazard is a piecewise constant function taking the values $r_{1,s_i}, \dots, r_{K, s_i}$ during time intervals $(t_0, t_1], \dots, (t_{K-1}, t_K]$. The relative baseline hazard during the reference interval is constant at $r_{\textref, s_i} = 1$.

In this reparameterization of the baseline hazard function, the relative baseline hazard is in fact equal to the relative probability a contact is infectious at calendar time $t$ compared to the reference interval: $r_{s_i}(t) = p_{c,0,s_i}(t)/p_{c, 0, s_i}(t_\textref)$. This allows us to use the epidemic curve as an informative prior for the relative baseline hazard parameters $r_{1, s_i}, \dots, r_{K, s_i}$. For site $s_i$, we specify independent priors $\log(r_{k,s_i}) \sim N(\mu_{k,s_i}, \sigma^2_{k,s_i})$ for $k = 1, \dots, K$, where $\mu_{k,s_i}$ is the epidemic curve's estimated log relative number of infections during interval $k$ compared to the reference interval, and $\sigma^2_{k,s_i}$ is derived from the confidence intervals that typically accompany epidemic curve data and may be scaled up to make the prior less informative.

For the reference interval, we recommend choosing the time interval with the most observed infections in the study in order to maximize the amount of information about $h_{\textref,s_i}$. We put a weakly informative prior on the baseline hazard during the reference interval so that the scale of the posterior baseline hazard function is driven by the observed rate of infections in the study. For site $s_i$, we specify the prior $\log(h_{\textref,s_i}) \sim N(\mu_{\textref, s_i}, \sigma^2_{\textref, s_i})$, where $\mu_{\textref, s_i}$ is the estimated log rate of infections per person-time during the reference interval, and $\sigma_{\textref, s_i}$ is derived from the epidemic curve's confidence intervals and may scaled up to make the prior less informative.

\subsection{Prior elicitation}

For the baseline hazard at each study site $s_i$, we use a weakly informative prior for $h_{\textref, s_i}$ and informative priors for the relative baseline hazard parameters $r_{1, s_i}, \dots, r_{K, s_i}$, as described in Section \ref{survival:sec:bl_hazard_prior}. For the relative infectiousness parameters $\alpha^\variant$ for each variant (if relevant), we recommend using existing data on circulating variants to form an informative prior in the form of $\alpha^\variant \sim N \big(\mu_\alpha^\variant, (\sigma^{\variant}_\alpha)^2 \big)$, where $\mu_\alpha^\variant$ is the estimated log relative infectiousness of variant $v$ compared to variant 1 and $\sigma^{\variant}_\alpha$ is the corresponding estimated SD. If estimating a threshold of protection $X_T$ for the predictor of interest, we recommend a uniform prior $X_T \sim \uniform[X_\ell, X_u]$, where $X_\ell$ and $X_u$ are chosen sensibly based on the observed range of values of the predictor of interest $X$. For the remaining parameters, we follow the recommendations of the Stan developers\supercite{stan_priors} and use weakly informative priors that are scaled to reflect the expected order of magnitude of the parameter, so that the prior does not place most of its mass on unrealistically extreme parameter values.\supercite{gelman_prior_llk, gelman_prior_vis} For real-valued coefficient parameters, we use $N(0,2)$ priors if we expect its absolute value to be above 0.1 or $N(0,0.5)$ priors if we expect its absolute value to be below 0.1 (this is relevant in our application to the MOMI-Vax study in Section \ref{survival:sec:momivax_analysis}, which includes several coefficients on the number of days before or after birth, which we expect to be very small). For multivariate coefficient parameters, such as $\bm{\beta}$, we use independent $N(0,2)$ or $N(0, 0.5)$ priors as described above for each element in the parameter vector.

\subsection{Implementation} \label{survival:sec:implementation}

For the simulations and data application in this paper, the posterior distributions were estimated using Stan\supercite{stan} via the interface RStan\supercite{rstan} in R.\supercite{R} When necessary, integration of the Normal PDF in the likelihood is approximated using Stan's \texttt{normal\_lcdf} function.\supercite{stan_normal_lcdf} Stan code for fitting this model is available in the Supporting Information. 

\section{Simulations} \label{survival:sec:simulations}

We applied our method to simulated COVID-19 vaccine trials with varying sample sizes and study designs. For each sample size ($N=$ 100, 500, 1,000, or 5,000), we simulated 50 COVID-19 vaccine trials with participants randomized 1:1 to either an active or placebo group. Enrollment dates were randomly sampled between July 1, 2021--December 1, 2021 for the vaccine group and between March 1--September 1, 2021 for the placebo group, so the vaccine group had an overall higher baseline hazard compared to the placebo group. Participants were randomly assigned to a site in the United States (US) state of Georgia, New York, or Washington, and were followed with visits every two months for up to six months or until they became infected. 

\subsection{Simulation methods}

For each participant $i$, we denoted their randomized intervention group $X_i$, and we generated an additional baseline covariate $Z_i \iid \bernoulli(0.5)$. We simulated infection times using the hazard function 
\begin{equation}
    h_i(t) = h_{\textref, s_i} r_{s_i}(t) \exp \Big\{ \gamma X_i + \beta Z_i \Big\}.
\end{equation}
To simplify our simulations, we set the number of circulating variants to 1. We randomly chose 20\% of infections to be interval censored, with lower and upper censoring dates set to the visit dates (at Months 0, 2, 4, or 6) before and after their true infection times. True parameter values were set to $\gamma = -1$ and $\beta = -0.5$. The baseline hazard function for each site was generated using the US state--level SARS-CoV-2 epidemic curve estimates from the Institute for Health Metrics and Evaluation (IHME).\supercite{ihme_data} 

For each simulated dataset, we fit our calendar-time survival model using one of the following priors for the relative baseline hazard parameters $\{r_{k, s_i}\}_{k=1}^K$ and the baseline hazard during the reference interval $h_{\textref, s_i}$ at each site. Figure \ref{survival:fig:bl_hazard_priors_sim} displays these prior distributions. 
\begin{enumerate}
    \item Correctly specified baseline hazard prior: The priors were formed using the IHME's SARS-CoV-2 epidemic curve data\supercite{ihme_data} following the method described in Section \ref{survival:sec:bl_hazard_prior}. For time interval $(t_{k-1}, t_k]$ and site $s_i$, the prior for the relative baseline hazard was $\log(r_{k,s_i}) \sim N(\mu_{k,s_i}, \sigma^2_{k,s_i})$, where the SD $\sigma_{k,s_i}$ was set to 0.25 (informative), 1 (moderately informative), 2.5 (weakly informative), or 5 (uninformative). For every site, the prior for the baseline hazard during the reference interval was $\log(h_{\textref, s_i}) \sim N(\mu_{\textref, s_i}, 5^2)$.
    \item Misspecified baseline hazard prior: The priors were formed using the IHME's SARS-CoV-2 epidemic curve data\supercite{ihme_data} from incorrect states (Pennsylvania, Missouri, or Ohio), 30 days prior. For a given time interval and site, the prior for the relative baseline hazard was $\log(r_{k,s_i}) \sim N(\mu_{k,s_i}, \sigma^2_{k,s_i})$, where the SD $\sigma_{k,s_i}$ was set to 0.25, 1, 2.5, or 5. For every site, the prior for the baseline hazard during the reference interval was $\log(h_{\textref, s_i}) \sim N(\mu_{\textref, s_i}, 5^2)$.
    \item Flat baseline hazard prior: For a given time interval and site, the prior for the relative baseline hazard was $\log(r_{k,s_i}) \sim N(\mu_{k,s_i}, \sigma^2_{k,s_i})$, where the SD $\sigma_{k,s_i}$ was set to 2.5 or 5. For every site, the prior for the baseline hazard during the reference interval was $\log(h_{\textref, s_i}) \sim N(-5, 5^2)$.
\end{enumerate}
We put $N(0,2)$ priors on $\gamma$ and $\beta$.

\subsection{Simulation results} \label{survival:sec:simulation_results}

\begin{figure}[H]
    
    \begin{center}
        \includegraphics[width=0.95\linewidth]{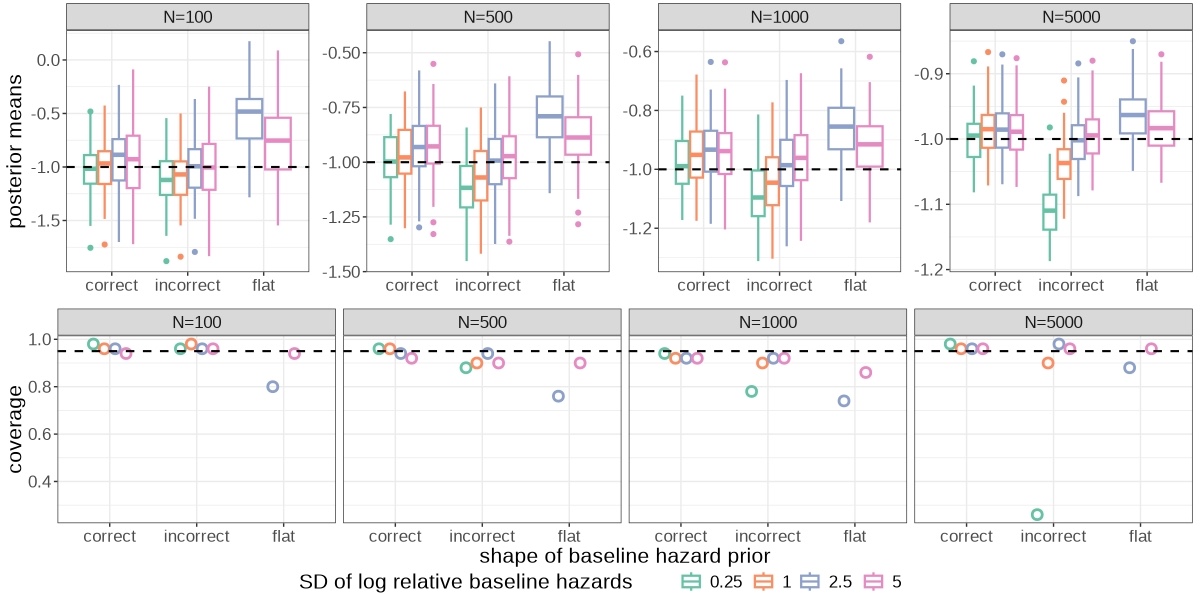}
    \end{center}

    \caption{Simulation results for the calendar-time survival model. The top and bottom rows show posterior mean estimates and 95\% posterior credible interval (CrI) coverage, respectively, for $\gamma$, the association between intervention group and the hazard of infection. Each column shows different simulated sample sizes. Different colors show different strengths of the baseline hazard prior used in fitting the model. The dashed lines shows the true value of $\gamma$ (above) and 95\% coverage (below).}
    \label{survival:fig:sim_results_gamma}
\end{figure}

Figure \ref{survival:fig:sim_results_gamma} shows the distribution of posterior means and the coverage of the 95\% posterior credible intervals (CrIs) for $\gamma$, the parameter of interest, which determines the association between intervention group and the hazard of infection. Table \ref{survival:table:sim_results} in the Supporting Information displays the full results for both $\gamma$ and $\beta$, which determines the association between the covariate $Z_i$ and the hazard of infection. When a correctly specified informative baseline hazard prior ($\sigma^2_{k,s_i}=0.25$) was used, then the posterior mean of $\gamma$ was unbiased and the 95\% CrI coverage was excellent, and correctly specified but less informative priors ($\sigma^2_{k,s_i}=1, 2.5,$ or $5$) also resulted in less biased estimates and improved coverage compared to the flat prior. However, if an a misspecified baseline hazard prior was used, then the posterior mean of $\gamma$ was more biased as it became more informative. The flat, uninformative prior (with $\sigma^2_{k,s_i}=5$) was only unbiased and had coverage about 95\% at a very large sample size ($N=5000$); in smaller sample sizes, the flat prior resulted in significant bias. For $\beta$, the coverage and bias were less sensitive to a misspecified informative baseline hazard prior.

\section{Application to the MOMI-Vax study} \label{survival:sec:momivax_analysis}

We now demonstrate the method introduced in this paper by applying it to data from the MOMI-Vax study, a multisite, prospective, longitudinal cohort study of mother-infant pairs following maternal COVID-19 vaccination during pregnancy. The primary aim of the study was to determine the association between several biomarkers (including anti--full-length spike [Spike] immunoglobulin G [IgG], pseudovirus neutralizing antibodies [NAbs] against the target vaccine strain, and live virus NAbs against several variants) in the infant and the infant's risk of SARS-CoV-2 infection within the first six months of life. In addition, the study aimed to compare infants whose mothers either received a complete series (2 doses) of a monovalent messenger RNA (mRNA) COVID-19 vaccine during pregnancy (the ``nonboosted" group) or who received a booster dose (3rd dose) during pregnancy (the ``boosted" group).\supercite{momivax_protocol} 

Cardemil et al\supercite{momivax_results} found that infants of mothers who had received a booster dose during pregnancy had significantly higher anti-Spike IgG titers, pseudovirus NAb titers, and live virus NAb titers at delivery and were 56\% less likely to be infected with SARS-Cov-2 during the first six months of life compared to infants of nonboosted mothers. When the SARS-CoV-2 case definition was expanded to include interval censored infections (identified through seroconversion of N protein between study visits), infants whose mothers had received a booster dose were 75\% less likely to be infected. These estimates were obtained using calendar time Cox proportional hazards models stratified by site, performed on a subset of the data from calendar times with sufficient follow-up in each dose group in order to ensure reliable inference.

As noted by Cardemil et al,\supercite{momivax_results} analysis of the MOMI-Vax data was complicated due to the timing of booster availability: the majority of the two-dose group of infants were born before the Omicron wave, while the majority of the three-dose group were born during the Omicron wave, leading to several months at the beginning and end of the study when only a single group was being followed. They addressed this problem by limiting the data to only periods of calendar time with $>$0.8 person-years of follow-up in both dose groups and in all infant age groups (0--2 months and 2--6 months). The method introduced in this paper enables estimation of the association between antibody levels at delivery and infection risk during the \textit{full} study period by using an informative prior for the baseline hazard of infection, allowing the model to borrow information from the prior during time periods with little or no data from a comparator group. Because we now retain data from the full study period, we now have many infections from the Delta period that the Cardemil et al analysis did not include. Our method allows us to estimate the association between antibodies and the risk of Delta or Omicron infection separately. Because the MOMI-Vax study did not collect information on which variant caused each infection, we use variant surveillance data from the US Centers for Disease Control and Prevention (CDC)\supercite{cdc_variant_data} to define the relative proportions of Delta and Omicron variants in circulation over time at each study site.

We now apply our method to estimate the association between several antibodies in the mother at delivery and the infant's risk of SARS-CoV-2 infection during the first six months of life. It is standard practice to collect blood from mothers when they are admitted for delivery,\supercite{msd_manual_labor} and maternal blood is typically easier and more reliable to collect than infants' cord blood.\supercite{acog_cord_banking} In addition, multiple studies,\supercite{maternal_antibody_transfer_review} including a preliminary analysis of the MOMI-Vax data,\supercite{momivax_antibodies} have shown that maternal SARS-CoV-2 antibodies (from both infection and vaccination) are transferred to the infant via the placenta, making them an ideal biomarker for assessing infants' protection against SARS-CoV-2 infection. 

Because a substantial proportion of mothers had antibody titers that had waned to below the LLOD by delivery, our model includes an association between antibody levels at delivery and infection risk \textit{only} for antibody levels above the LLOD. For maternal antibodies below the LLOD at delivery, we assume the hazard of infection is equal to that of mothers who have \textit{no} antibodies at delivery, conditional on the covariates included in $\bm{Z}_i(t)$. We also fit a second model in which we estimate the maternal antibody threshold below which infants' hazard of infection is equal to that of mothers with no antibodies. This second analysis gives insight into the minimum level at which maternal antibodies protect infants against SARS-CoV-2 infection.

\subsection{MOMI-Vax analysis methods} \label{survival:sec:momivax_model}

We assessed the relationship between three antibodies in the mothers at delivery and their infants' risk of SARS-CoV-2 infection: anti-Spike IgG, pseudovirus NAbs against the ancestral Wuhan strain (614BD (expressed as the concentration required to inhibit 50\% of viral entry and replication [IC50]), and live virus NAbs to the target vaccine strain (D614G) (expressed as the inhibitory dilution required to achieve 50\% neutralization [ID50]). For brevity, we refer to these antibodies as anti-Spike IgG, pseudovirus NAbs, and target strain live virus NAbs. For each antibody of interest, we performed two analyses: first, an analysis using the LLOD as the threshold beneath which the antibodies have no association with the hazard of infection, and second, an analysis in which we estimate this threshold as a parameter itself. The outcome was defined as SARS-CoV-2 infection identified either through verified maternal report (confirmed by laboratory testing) or through seroconversion of N protein between study visits.

The hazard functions for each model are shown below. The superscripts $(d)$ and $(o)$ denote the Delta and Omicron variant, respectively; $X_i$ is the mother's $\log_{10}$ antibody level at delivery, $X_\llod$ is the $\log_{10}$ LLOD for the antibody of interest; $X_T$ is the estimated $\log_{10}$ antibody threshold (for the estimated threshold analysis); and $\bm{Z}_i$ is a vector of baseline covariates. To match the Cardemil et al analysis,\supercite{momivax_results} $\bm{Z}_i$ includes the following baseline covariates: the mother's primary COVID-19 vaccine series type (Moderna or Pfizer), race, ethnicity, the number of maternal comorbidities, maternal body mass index (BMI), the mother being a healthcare worker, the mother's Occupational Safety and Health Administration (OSHA) exposure risk, and the mother's work from home status. Data for $\pi^\dvar_{s_i}(t)$ and $ \pi^\ovar_{s_i}(t)$ came from the CDC's SARS-CoV-2 variant surveillance data.\supercite{cdc_variant_data}  The infecting variants were not collected in the MOMI-Vax study, so when fitting the model, we use the form of the likelihood for unknown infecting variants presented in Section \ref{survival:sec:llk_unknown_variants}.

The hazard function for the LLOD threshold analysis is
    \begin{equation}
    \begin{aligned}
        h_i(t) &= \pi^\dvar_{s_i}(t) h_{\textref, s_i} r_{s_i}(t) \exp \Big\{ \gamma^\dvar X_i  \ind\{ X_i > X_{\llod} \} + \gamma_\llod^\dvar \ind\{ X_i > X_{\llod} \} + \bm{Z}_i \bm{\beta}_h \Big\} \\
        &\hphantom{=} + \pi^\ovar_{s_i}(t) h_{\textref, s_i} r_{s_i}(t) \exp \Big\{ \alpha^\ovar + \gamma^\ovar X_i \ind\{ X_i > X_{\llod} \} + \gamma_\llod^\ovar \ind\{ X_i > X_{\llod} \} + \bm{Z}_i \bm{\beta}_h \Big\}.
    \end{aligned}
    \end{equation}

The hazard function for the estimated threshold analysis is
    \begin{equation}
    \begin{aligned}
        h_i(t) &= \pi^\dvar_{s_i}(t) h_{\textref, s_i} r_{s_i}(t) \exp \Big\{ \gamma^\dvar X_i  \ind\{ X_i > X_T \} + \gamma_T^\dvar \ind\{ X_i > X_T \} + \bm{Z}_i \bm{\beta}_h \Big\} \\
        &\hphantom{=} + \pi^\ovar_{s_i}(t) h_{\textref, s_i} r_{s_i}(t) \exp \Big\{ \alpha^\ovar + \gamma^\ovar X_i \ind\{ X_i > X_T \} + \gamma_T^\ovar \ind\{ X_i > X_T \} + \bm{Z}_i \bm{\beta}_h \Big\}.
    \end{aligned}
    \end{equation}
For the estimated threshold analysis, $X_T$ is given the prior $X_T \sim \text{Unif}(X_\llod, X_{1/2})$, where $X_{1/2}$ is the median observed biomarker value among those above the LLOD.

\begin{figure}[H]
    
    \begin{center}
        \includegraphics[width=1\linewidth]{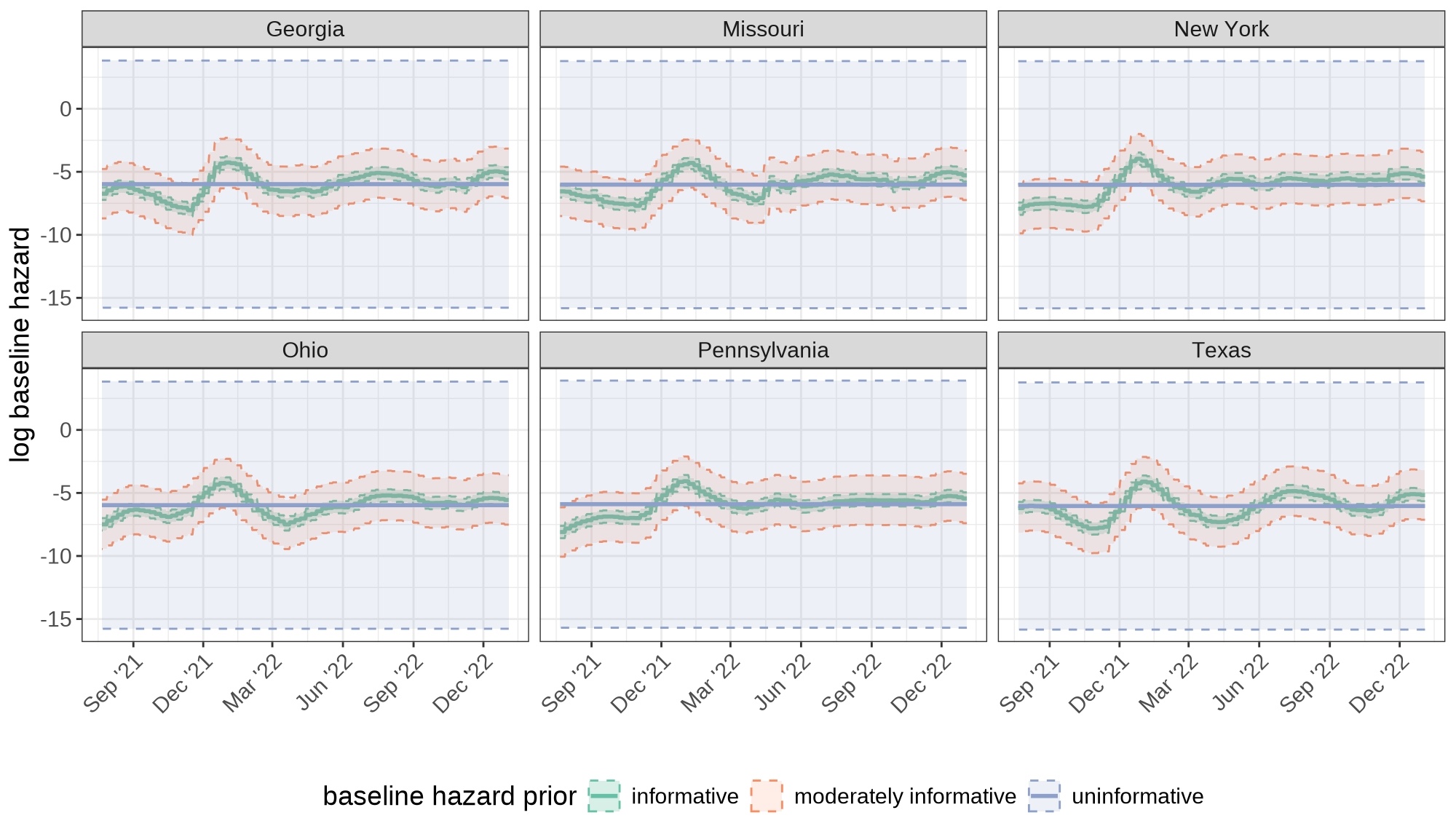}
    \end{center}

    \caption{Priors for the baseline hazard function used in the MOMI-Vax analysis. The solid lines show the means of each prior distribution, while the ribbons show the 95\% quantiles of each prior distribution. The value of $h_{\textref, s_i}$ (the baseline hazard during the reference interval of calendar time) is fixed at its mean in these plots in order to highlight the level of informativeness of the relative baseline hazard priors.}
    
    \label{survival:fig:bl_hazard_priors_momivax}
\end{figure}

For each antibody and threshold analysis, we fit our model using three different priors for the relative baseline hazard function parameters $\{r_{k, s_i}\}_{k=1}^K$. We used the IHME's SARS-CoV-2 epidemic curve estimates\supercite{ihme_data} to form the informative and moderately informative baseline hazard priors following the method described in Section \ref{survival:sec:bl_hazard_prior}. Figure \ref{survival:fig:bl_hazard_priors_momivax} shows the priors used in the MOMI-Vax analyses.
\begin{enumerate}
    \item Informative baseline hazard prior: For each calendar time interval $(t_{k-1}, t_k]$ and site $s_i$, $\log(r_{k, s_i}) \sim N(\mu_{k, s_i}, 0.25^2)$ and $\log(h_{\textref, s_i}) \sim N(\mu_{\textref, s_i}, 5^2)$.
    \item Moderately informative baseline hazard prior: For each calendar time interval $(t_{k-1}, t_k]$ at each site $s_i$, $\log(r_{k, s_i}) \sim N \big( \mu_{k, s_i}, 1 \big)$ and $\log(h_{\textref, s_i}) \sim N(\mu_{\textref, s_i}, 5^2)$.
    \item Uninformative baseline hazard prior: For all time intervals $(t_{k-1}, t_k]$ at each site $s_i$, the mean of the relative baseline hazards was set to $\overline{\mu}_{k, s_i}$, the mean relative baseline hazard reported by the IHME in that site over all time intervals, and the SDs were all set to 2.5. The resulting priors were $\log(r_{k, s_i}) \sim N( \overline{\mu}_{k, s_i}, 5^2)$ and $\log(h_{\textref, s_i}) \sim N(\mu_{\textref, s_i}, 5^2)$.
\end{enumerate}
For $\alpha^\ovar$, we used the informative prior $\alpha^\ovar \sim N \big( \log(2), 0.2 \big)$ based on previous research suggesting that the Omicron variant is approximately twice as infectious as the Delta variant.\supercite{delta_vs_omicron_transmission} For $\gamma^\dvar$ and $\gamma^\ovar$ in the estimated threshold analysis, we used a truncated Normal prior $\gamma^\dvar, \gamma^\ovar \iid N(0, 2)\ind(-\infty, 0)$, which constrains the maternal antibody levels to be associated with only a reduced risk of infection compared to levels below the threshold. In the LLOD threshold analysis, we used the prior $\gamma^\dvar, \gamma^\ovar \iid N(0, 2)$. For the remaining parameters $\gamma_\llod^\dvar$, $\gamma_\llod^\ovar$, $\gamma_T^\dvar$, $\gamma_T^\ovar$, and $\bm{\beta}$, we used $N(0,2)$ priors.

For each antibody of interest, threshold analysis, and baseline hazard prior, we fit a Stan model with 4 chains of 4000 iterations (for the LLOD threshold analyses) or 8000 iterations (for the estimated threshold analyses), discarding the first half of each chain as burn-in. To assess convergence, we evaluated the posterior trace plots and ensured $\hat{R}$ was below 1.05 for all parameters.\supercite{rhat} To assess model fit, we randomly selected 100 samples from the posterior and generated predicted survival times for each infant. Because a significant amount of infections were interval censored between Months 0, 2, or 6, we compared the predicted cumulative incidences to the observed cumulative incidences at Months 2 and 6 to assess model fit.

\subsection{Results of the MOMI-Vax analysis} \label{survival:sec:momivax_results}

\begin{figure}[H]

    \begin{center}
        \includegraphics[width=0.85\linewidth]{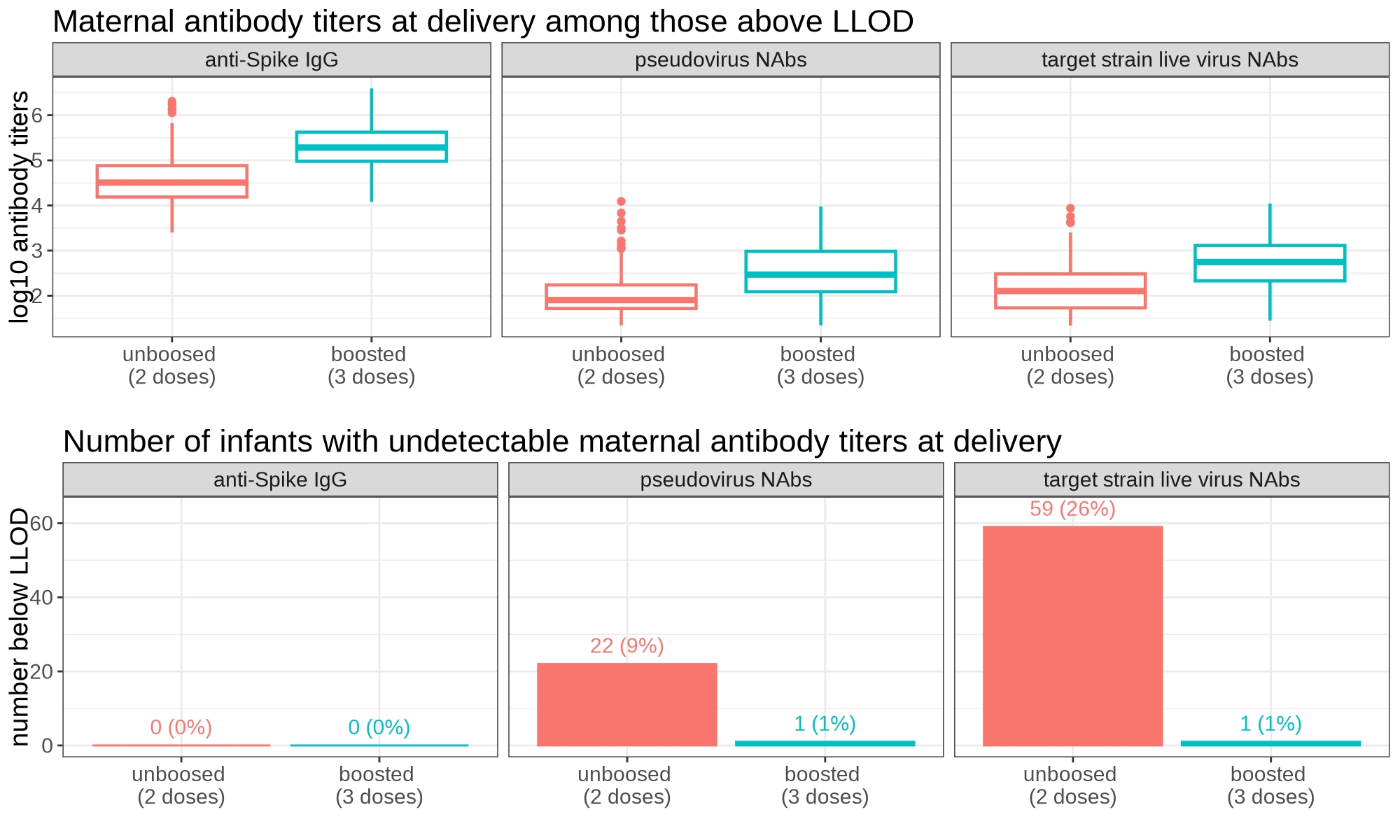}
    \end{center}
    
    \caption{Observed maternal antibodies at delivery in the MOMI-Vax study. The top row shows the distribution of maternal antibody titers among those above the lower limit of detection (LLOD) by dose group, while the bottom plot shows the number and proportion of infants in each dose group whose mothers had undetectable antibodies (below the LLOD) at delivery.}
    
    \label{survival:fig:maternal_antibodies}
\end{figure}

Table \ref{survival:table:momivax_table1} summarizes the baseline characteristics and SARS-CoV-2 infection outcomes among infants in the MOMI-Vax study. Our analysis included 457 infants whose mothers either received a complete series (2 doses) of a monovalent messenger RNA (mRNA) COVID-19 vaccine during pregnancy or who received a booster dose (3rd dose) during pregnancy, whose mothers had at least one antibody measurement available at delivery, and who had complete data for all of the baseline covariates in our model. Of these infants, 263 were born to nonboosted (2 dose) mothers and 194 were born to boosted (3 dose) mothers. The nonboosted group was born between July 2021 and June 2022 (median October 2021), while the boosted group was born between October 2021 and August 2022 (median February 2022). During the first 6 months of life, 100 infants (38\%) of nonboosted mothers were infected with SARS-CoV-2, and 86 infants (44\%) of boosted mothers were infected with SARS-CoV-2. Only 4 uncensored infections occurred during the Delta period (before mid-to-late December 2021, depending on the state, when the majority of circulating variants were Delta), and 51 interval censored infections might have occurred as early as the Delta period, most of which also spanned the Omicron period. For all antibodies considered in this analysis, boosted mothers had higher median titers at delivery than nonboosted mothers (Figure \ref{survival:fig:maternal_antibodies}). All anti-Spike IgG measurements were in the detectable range (above the LLOD) at delivery, while 22 nonboosted mothers and 1 boosted mother had undetectable pseudovirus NAb measurements at delivery, and 64 nonboosted mothers and 1 boosted mother had undetectable live virus NAb measurements at delivery. 

\begin{figure}[H]

    \begin{center}
        \includegraphics[width=0.8\linewidth]{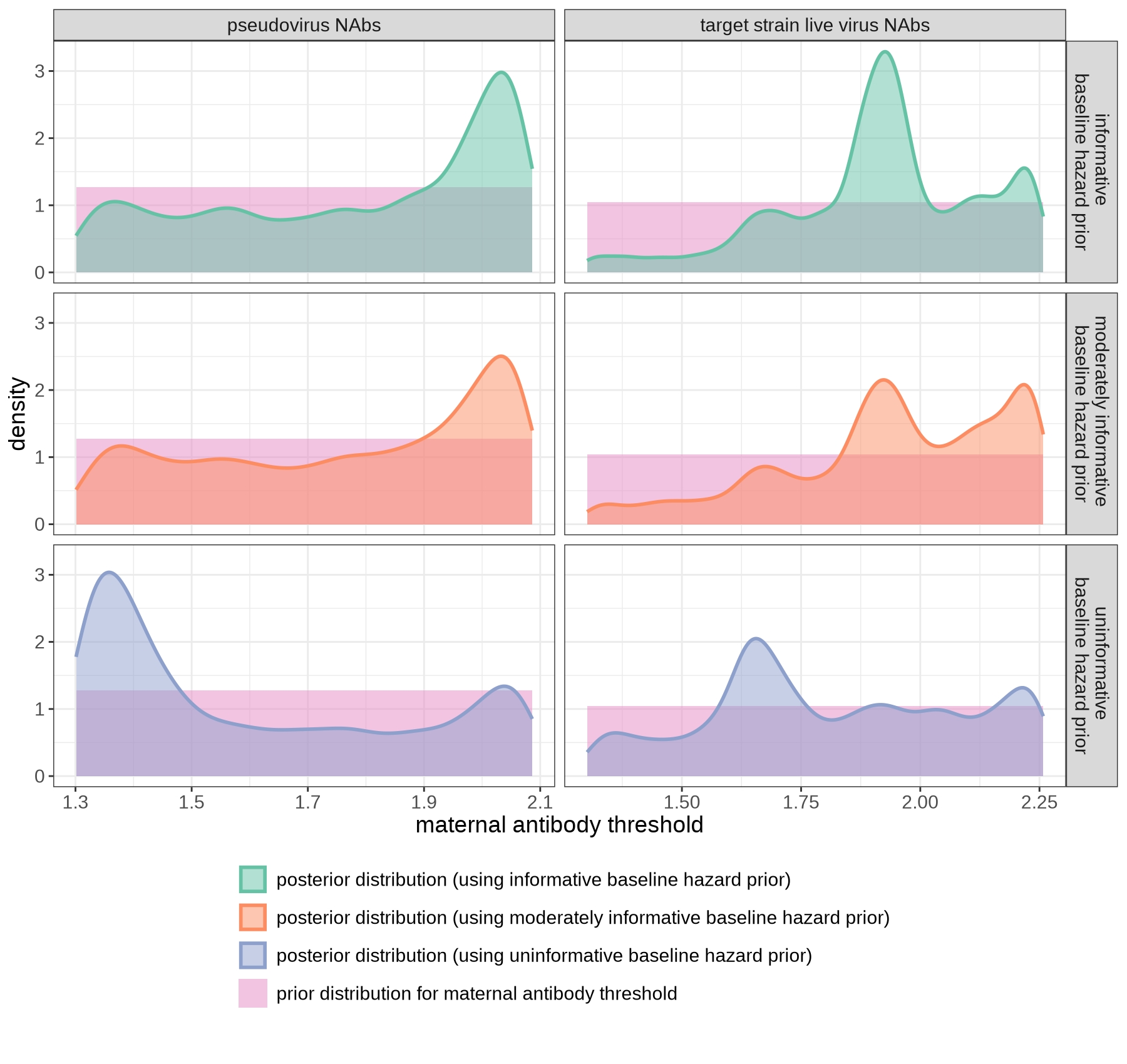}
    \end{center}

    \caption{Prior and posterior distributions for the threshold at which maternal antibody levels at delivery are associated with infants' risk of SARS-CoV-2 infection in the MOMI-Vax study. Below the threshold, the model assumes that the antibodies have no association with the risk of infection.}
    
    \label{survival:fig:X_T}
\end{figure}

Table \ref{survival:table:momivax_gamma_results} displays the posterior means and 95\% CrIs for the parameters of interest related to the association between maternal antibodies and the risk of SARS-CoV-2 infection from either the Delta or Omicron variant, as well as the estimated threshold at which maternal antibodies are associated with infants' infection risk. Figure \ref{survival:fig:X_T} summarizes the posterior distributions for this threshold. The model was unable to identify an anti-Spike IgG threshold. Because so few infants were infected during the Delta period, most of whom were in the nonboosted group, the estimates for the Delta period are unreliable; therefore, only the Omicron results are included in the figures in this paper. Tables and figures containing the results for the Delta period are available in the Supporting Information (see Figures \ref{survival:fig:gamma_delta} and \ref{survival:fig:gamma_3vs2_delta}). 

\begin{figure}[H]

    \begin{center}
        \includegraphics[width=0.8\linewidth]{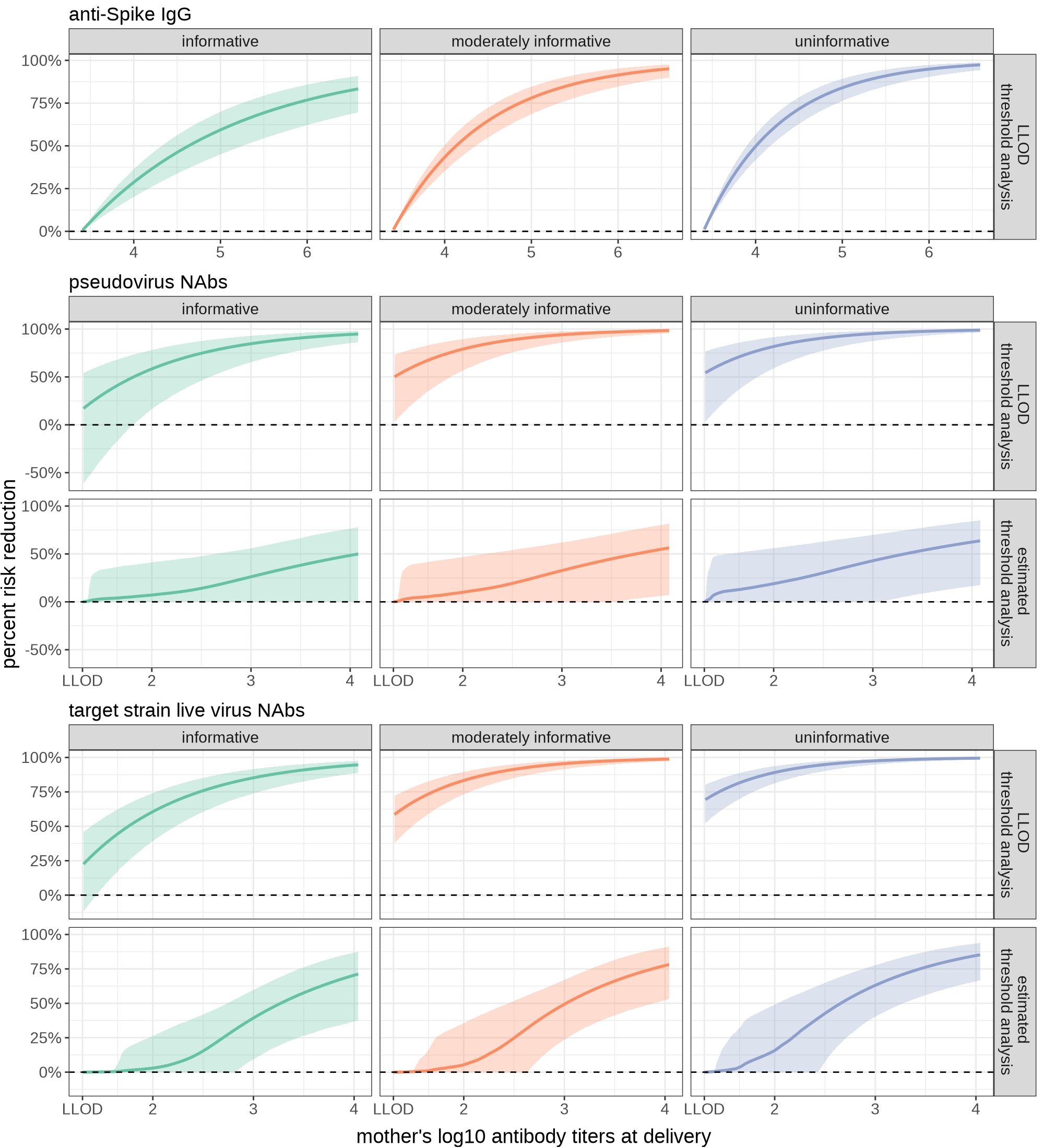}
    \end{center}

    \caption{Estimated association between maternal antibodies at delivery and infants' risk of SARS-CoV-2 infection with the Omicron variant in the MOMI-Vax study. Posterior means and 95\% posterior credible intervals (CrIs) are shown for the percent reduction in risk of infection in infants compared to the minimum observed maternal antibody titers in the study (top row for each antibody) or compared to the estimated maternal antibody threshold (bottom row for each antibody), adjusted for the covariates included in $\bm{Z}$. All comparisons are among infants at the same site and at the same calendar time. Relative risk reduction is calculated as (1 -- hazard ratio) and is only shown for the observed range of maternal antibodies in the study. The faceting columns refer to the baseline hazard prior used in the model.}
    
    \label{survival:fig:gamma_omicron}
\end{figure}

\begin{figure}[H]
    \begin{center}
        \includegraphics[width=1\linewidth]{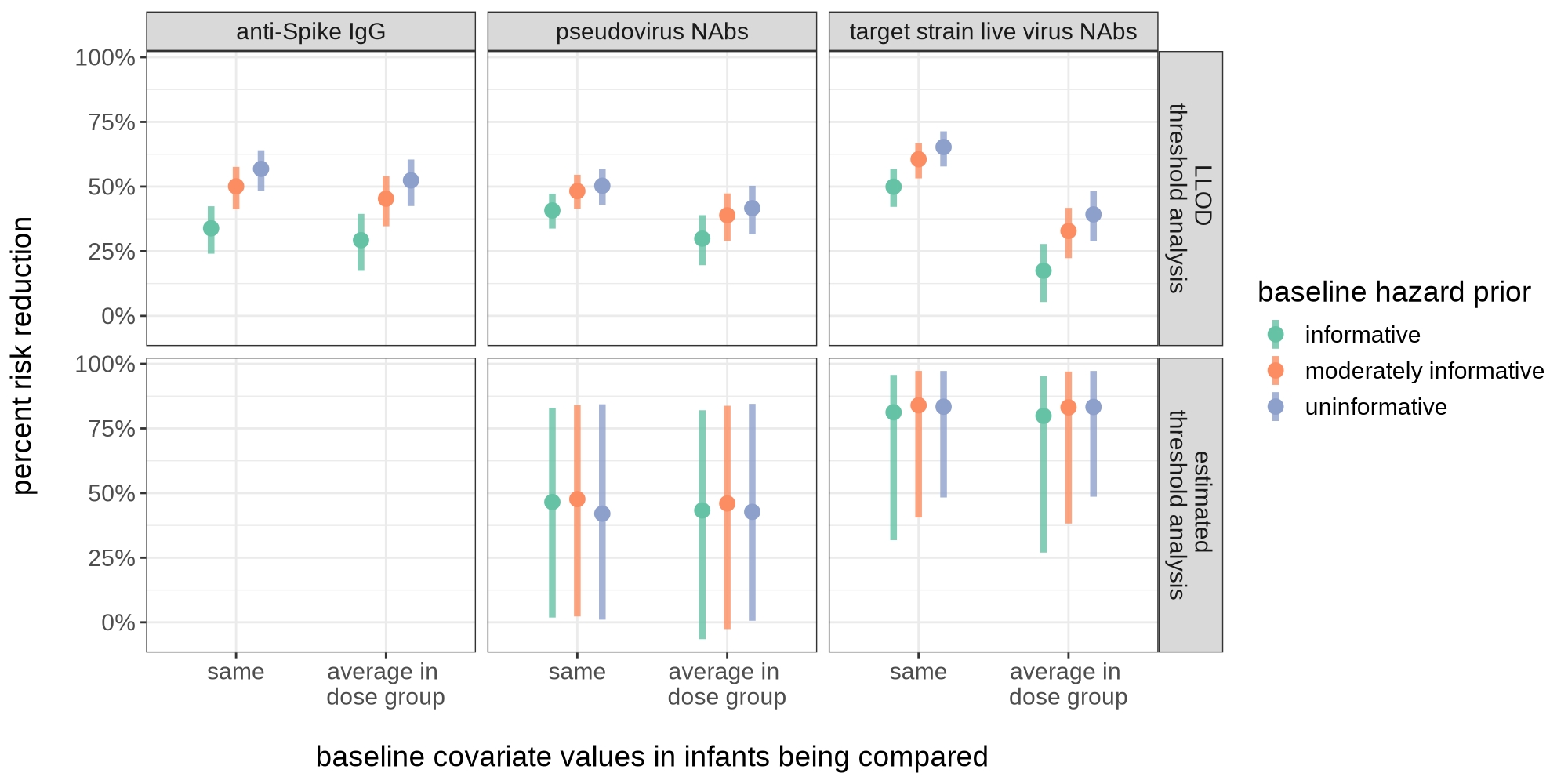}
    \end{center}

    \caption{Estimated relative risk of infection with the SARS-CoV-2 Omicron variant among average boosted versus unboosted infants in the MOMI-Vax study. Posterior means and 95\% posterior credible intervals (CrIs) are shown for the percent reduction in risk of infection with the SARS-CoV-2 Omicron variant in infants in the MOMI-Vax study, comparing infants whose mothers had the average boosted (3 dose) antibody levels vs. the average nonboosted (2 dose) antibody levels at delivery. Relative risk reduction is calculated as (1 -- hazard ratio). For each antibody and baseline hazard prior, results are shown for two comparisons: infants with the same baseline covariates who differ only in their maternal antibody levels (left); and infants whose baseline covariates are set to the average values in their dose group (right), highlighting the differences in risk between the 3 and 2 dose groups beyond calendar time and maternal antibody levels. All comparisons are among infants at the same site and at the same calendar time.}
    \label{survival:fig:gamma_3vs2_omicron}
\end{figure}

All analyses found that higher maternal antibody titers were associated with a reduced risk of SARS-CoV-2 infection with the Omicron variant; however, the informative baseline hazard prior attenuated the association toward less risk reduction, suggesting that model with the uninformative baseline hazard prior overestimates the association between maternal antibodies and reduced risk of infection. To aid in interpreting the results, Figure \ref{survival:fig:gamma_omicron} shows the posterior relative risk reduction for SARS-CoV-2 infection in infants at different maternal antibody levels compared to either the minimum observed antibody level in the study or the estimated antibody threshold. In addition, to highlight the differences in risk between the nonboosted and boosted groups beyond calendar time and maternal antibody levels, Figure \ref{survival:fig:gamma_3vs2_omicron} shows the posterior relative risk reduction for SARS-CoV-2 infection in infants, comparing ``average" boosted infants to ``average" nonboosted infants.

The posterior distributions for the association between the baseline covariates and infection risk were similar across the analyses using different baseline hazard priors and different antibodies (see Table \ref{survival:table:momivax_results_beta_h} and Figure \ref{survival:fig:beta_h} in the Supporting Information). All Stan models converged to a posterior with MCMC chains that mixed well with $\hat{R}<1.05$ for all parameters, and all models predict cumulative incidence of SARS-CoV-2 infection in the infants very close to the observed cumulative incidence (see Figures \ref{survival:fig:traceplots} and  \ref{survival:fig:cuminc} in the Supporting Information).

\section{Discussion} \label{survival:sec:discussion}

We have developed a Bayesian survival model in calendar time that is tailored to infectious disease studies occurring during an epidemic. We introduced a method for creating an informative prior for the baseline hazard function, enabling the incorporation of outside data about the epidemic into our understanding of participants' risk of exposure to infection during the study. We formulated our model as a competing risks model, which allows us to estimate variant-specific associations between the predictor of interest and the risk of infection. For studies in which a biomarker is the predictor of interest, we included the option to estimate a threshold of protection for the biomarker. Because our method has a fully specified likelihood, it easily incorporates interval censoring in addition to right and left censoring.

Our simulation results demonstrate that when calendar time bias causes the comparator groups to have very different baseline hazards of infection, using an informative prior for the shape of the baseline hazard function corrects the bias in estimates of $\gamma$, the association between the predictor of interest and the hazard of infection, that is otherwise present when an uninformative prior is used. However, using a misspecified informative prior can induce bias in estimates of both $\gamma$ and other coefficients in the hazard function, emphasizing the need for careful consideration of whether the epidemic curve used to create the informative prior truly reflects the changing risk of exposure in the study population. State-level epidemic curve data such as the IHME data used in this paper may not sufficiently reflect the relative risk of exposure in certain studies, such as studies in rural populations or with targeted enrollment of high-risk participants.

When we applied our method to the MOMI-Vax study, we found that higher maternal antibody titers at delivery were associated with reduced risk of infection in their infants during the first six months of life. Using an informative baseline hazard prior attenuated this association toward less protection against infection. This attenuation is unsurprising: the infants of nonboosted mothers were largely born at a low point in the epidemic, so their antibody levels were highest when infection risk was comparatively low, while the infants of boosted mothers were largely born at higher points in the epidemic, during the first or second Omicron waves. Without an informative baseline hazard prior, the model likely attributes the lack of early infections in the nonboosted group to an especially strong antibody effect, as nonboosted mothers' antibody levels were much lower than boosted mothers'. With an informative baseline hazard prior, the model knows that the nonboosted group had a lower baseline risk of infection during this time, decreasing the strength of the antibodies' association with infection risk. 

We also applied our method to the MOMI-Vax data to estimate a threshold above which maternal pseudovirus and target strain live virus antibodies are associated with protection from SARS-CoV-2 infection in their infants. This is the first time a threshold at which antibodies provide protection against SARS-CoV-2 infection has been estimated for infants. When we incorporated outside information about the epidemic through an informative baseline hazard prior, our results suggest the need for slightly higher maternal antibody levels to provide their infants with protection from SARS-COV-2 infection during the first six months of life, which a booster dose during pregnancy helps provide. It should be noted that the estimated threshold should not be interpreted as a causal or biological cutoff, but rather an approximate point at which antibodies are associated with protection from infection in this study population and under this model. Because our results are from an observational study, more research is needed, preferably from a randomized study, to establish a threshold of protection for infants. Future research may also consider different relationships between the antibodies and infection risk, such as a logistic relationship, rather than a linear relationship with a sharp cutoff at the threshold.

Ideally, we would have liked to use the method proposed in this paper to perform a mediation analysis to quantify how much infants' protection against SARS-CoV-2 infection is mediated through their or their mothers' antibodies. However, the observational nature of the MOMI-Vax data made calculating counterfactual outcomes impossible: many nonboosted mothers received their second vaccine dose during the third trimester of pregnancy, making a booster dose impossible, which violates the positivity assumption required for a causal mediation analysis.\supercite{mediation_analysis_assumptions} More research will be needed to establish a causal association between infant or maternal antibodies and infants' risk of SARS-CoV-2 infection.

Our results are in line with previous work on immune correlates of protection against COVID-19 disease in vaccinated adults. While the exact estimates vary by study, anti-Spike IgG titers, pseudovirus NAbs, and (if measured) live virus NAbs been shown to reliably predict vaccine-elicited protection against COVID-19.\supercite{moderna_correlates, novavax_correlates, janssen_correlates, covid_correlates_summary} Along with the previously published MOMI-Vax results for infants' antibodies at birth,\supercite{momivax_results} our results support these antibodies as predictors of protection against SARS-CoV-2 infection in infants. However, we caution that the MOMI-Vax results are not directly comparable to previously published results on correlates of protection against SARS-CoV-2 infection for several reasons. First, the MOMI-Vax study was observational, and the nonboosted and boosted groups different significantly in their characteristics; for example, mothers who received a booster dose during pregnancy were more likely to be nonwhite and to have a higher risk of exposure at work. There are very likely additional, unmeasured confounders affecting the association between maternal antibodies and infants' infection risk in the MOMI-Vax study. Second, the MOMI-Vax study used SARS-CoV-2 infection as its outcome, rather than symptomatic COVID-19 disease, as was used in most COVID-19 vaccine efficacy trials. Third, the study population (infants whose mothers had received a COVID-19 vaccine) is very different from the adult populations on which the previous results are based on. Finally, the MOMI-Vax study did not include any mothers who had never been vaccinated, unlike randomized trials that included a placebo group. 

Our method for incorporating epidemic curve data into an informative prior for the baseline hazard of infection is applicable to a variety of settings. In particular, household transmission studies could benefit from epidemic curve data to inform the risk of infection from the community; such studies are typically case-ascertained and lack information about sources of infection outside the household, which can lead to biased estimates of the rate of transmission from household contacts.\supercite{klick2014optimal-4db}


\section*{Acknowledgements}

MOMI-Vax study data were provided by the Infectious Diseases Clinical Research Consortium (IDCRC) through the National Institute of Allergy and Infectious Diseases (NIAID), part of the National Institutes of Health (NIH), under award number UM1AI148684. The content is solely the responsibility of the authors and does not necessarily represent the official views of the IDCRC or the NIH.\vspace*{-8pt}

\section*{Supporting Information}

Supplementary figures referenced in Sections \ref{survival:sec:simulation_results} and \ref{survival:sec:momivax_results} as well as Stan code for fitting the model as referenced in Section \ref{survival:sec:implementation} are available in the Supporting Information.\vspace*{-8pt}

\printbibliography

\newpage

\section*{Tables and figures}

\begin{Table}[H]
        
    \caption{Baseline characteristics and infection outcomes among infants in the MOMI-Vax study.}
    
    \vspace{0.1in}

    \begin{center}
        \includegraphics[width=1\linewidth]{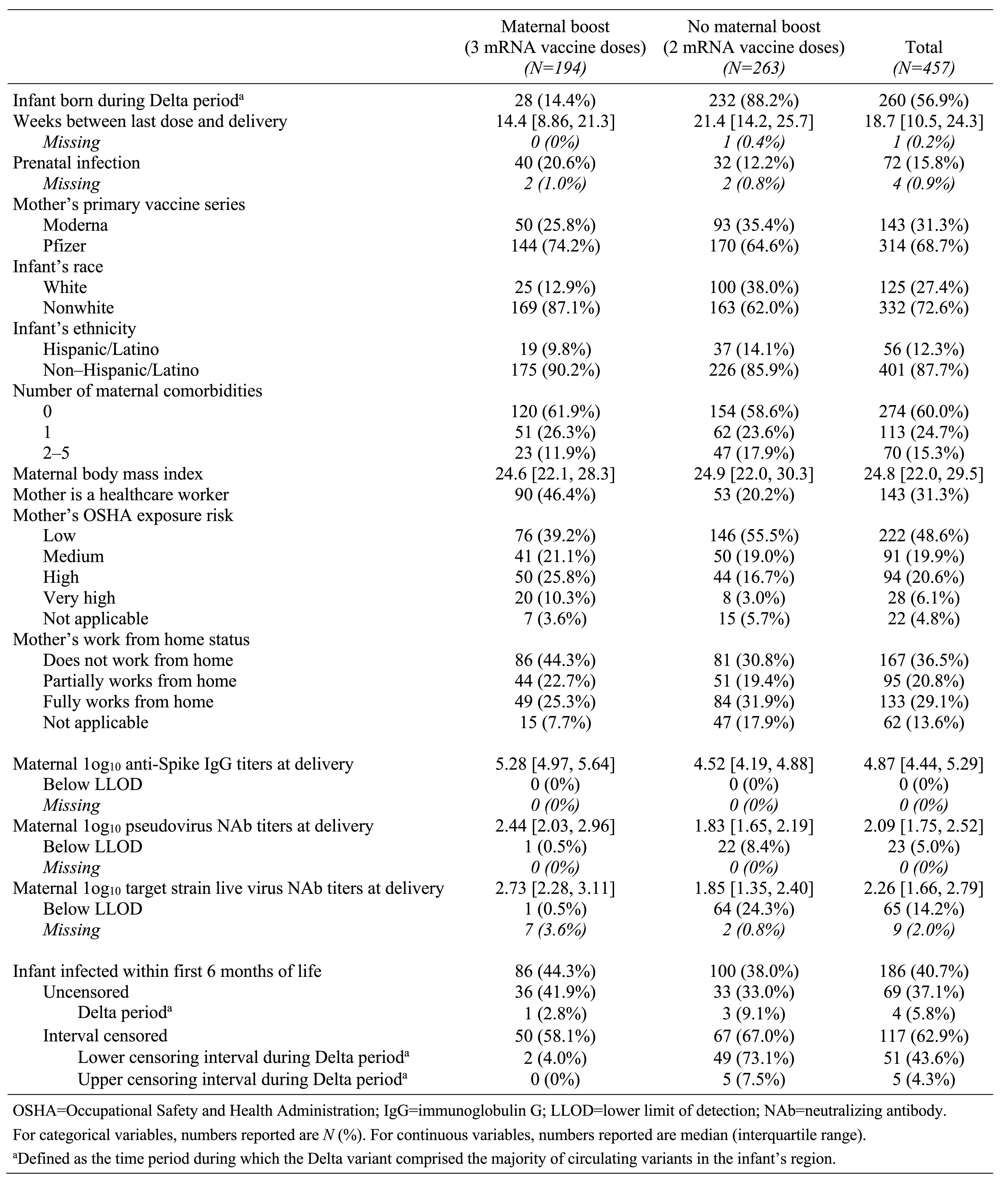}
    \end{center}
    
    \label{survival:table:momivax_table1}
\end{Table}

\begin{landscape}

\begin{Table}[H]
    
    \caption{Primary results for the MOMI-Vax data analysis using the calendar-time survival model.} 
    

    \begin{center}
        \includegraphics[width=0.74\linewidth]{tables_and_listings/momivax_gamma_results_table_v2.jpeg}
    \end{center}

    Results are shown for two analyses: one using the lower limit of detection (LLOD) as the threshold below which the model assumes no association between maternal antibodies and the risk of infection, and a second in which this threshold is estimated. Relative risk reduction (RRR) is calculated as (1 -- hazard ratio). Positive RRR means the predictor is associated with a reduced risk of infection, while negative RRR means the predictor is associated with an increased risk of infection.
    
    \label{survival:table:momivax_gamma_results}
\end{Table}

\end{landscape}

\clearpage
\section*{Supporting Information}
\setcounter{page}{1}
\renewcommand{\thepage}{Supporting information page \arabic{page}}

\setcounter{figure}{0}
\renewcommand{\thefigure}{S\arabic{figure}}

\setcounter{Table}{0}
\renewcommand{\theTable}{S\arabic{Table}}

\begin{figure}[H]

    \begin{center}
        \includegraphics[width=\linewidth]{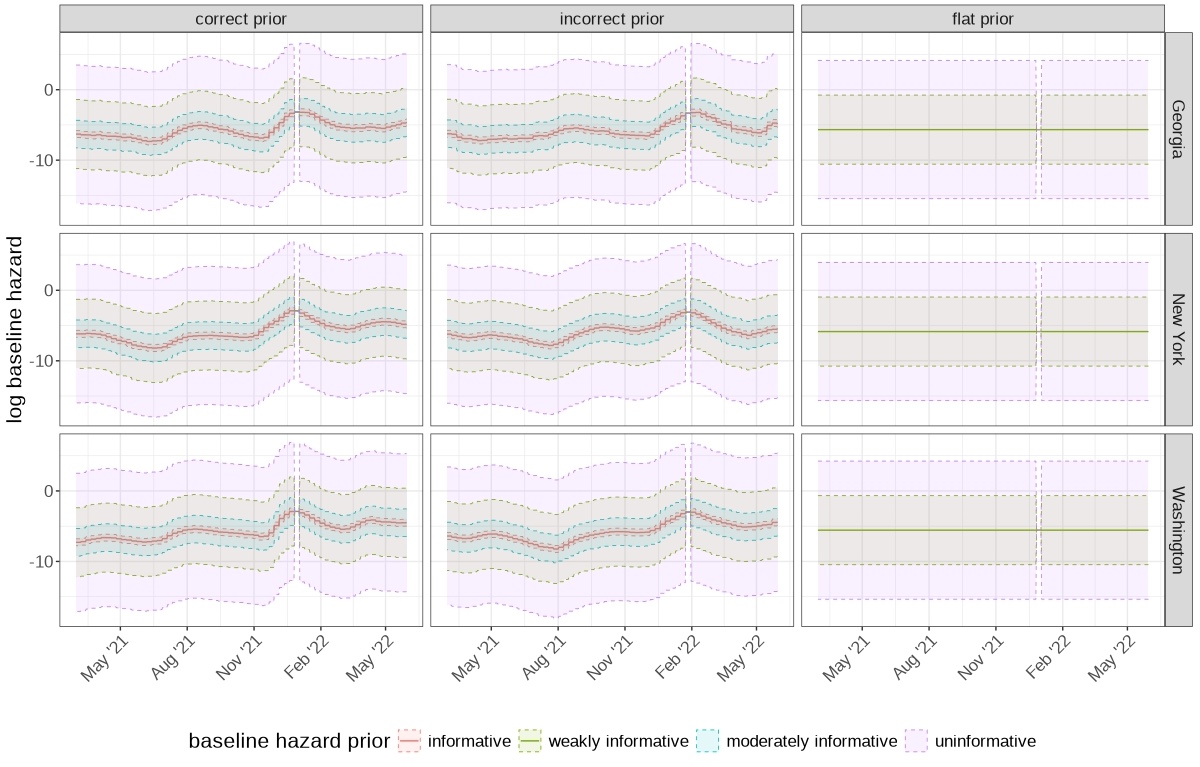}
    \end{center}
    
    \caption{Priors for the baseline hazard function used in simulations. The solid lines show the means for each prior distribution, while the ribbons show the 95\% quantiles. The value of $h_{\textref, s_i}$ is fixed at its mean in these plots to show the level of informativeness of the relative baseline hazard priors.}
    \label{survival:fig:bl_hazard_priors_sim}
\end{figure}

\begin{Table}[H]
        
    \caption{95\% posterior credible interval (CrI) coverage and bias of posterior means from calendar-time survival model simulations.} 

    \begin{center}
        \includegraphics[width=0.75\linewidth]{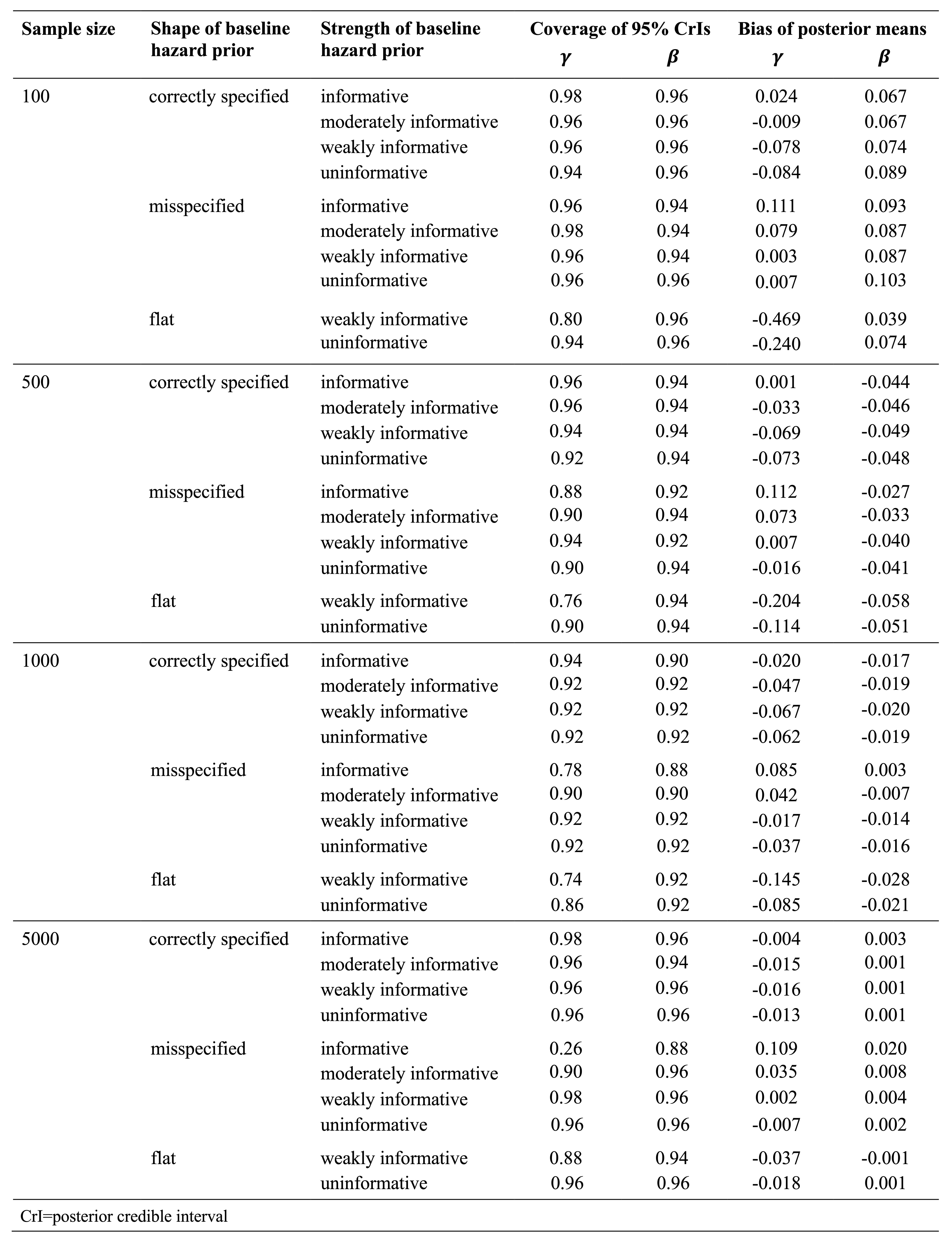}
    \end{center}

    Results are shown for $\gamma$, the association between intervention group (vaccine vs. placebo) and the risk of infection (adjusted for the baseline covariate $Z_i$), and $\beta$, the association between $Z_i$ and the risk of infection (adjusted for intervention group). The strength of the baseline hazard priors refer to different standard deviations of the priors for the log relative baseline hazards: informative ($\sigma_{k, s_i}=0.25$), moderately informative ($\sigma_{k, s_i}=1$), weakly informative ($\sigma_{k, s_i}=2.5$), or uninformative ($\sigma_{k, s_i}=5$).

    \label{survival:table:sim_results}
\end{Table}

\begin{figure}[H]

    \begin{center}
        \includegraphics[width=0.9\linewidth]{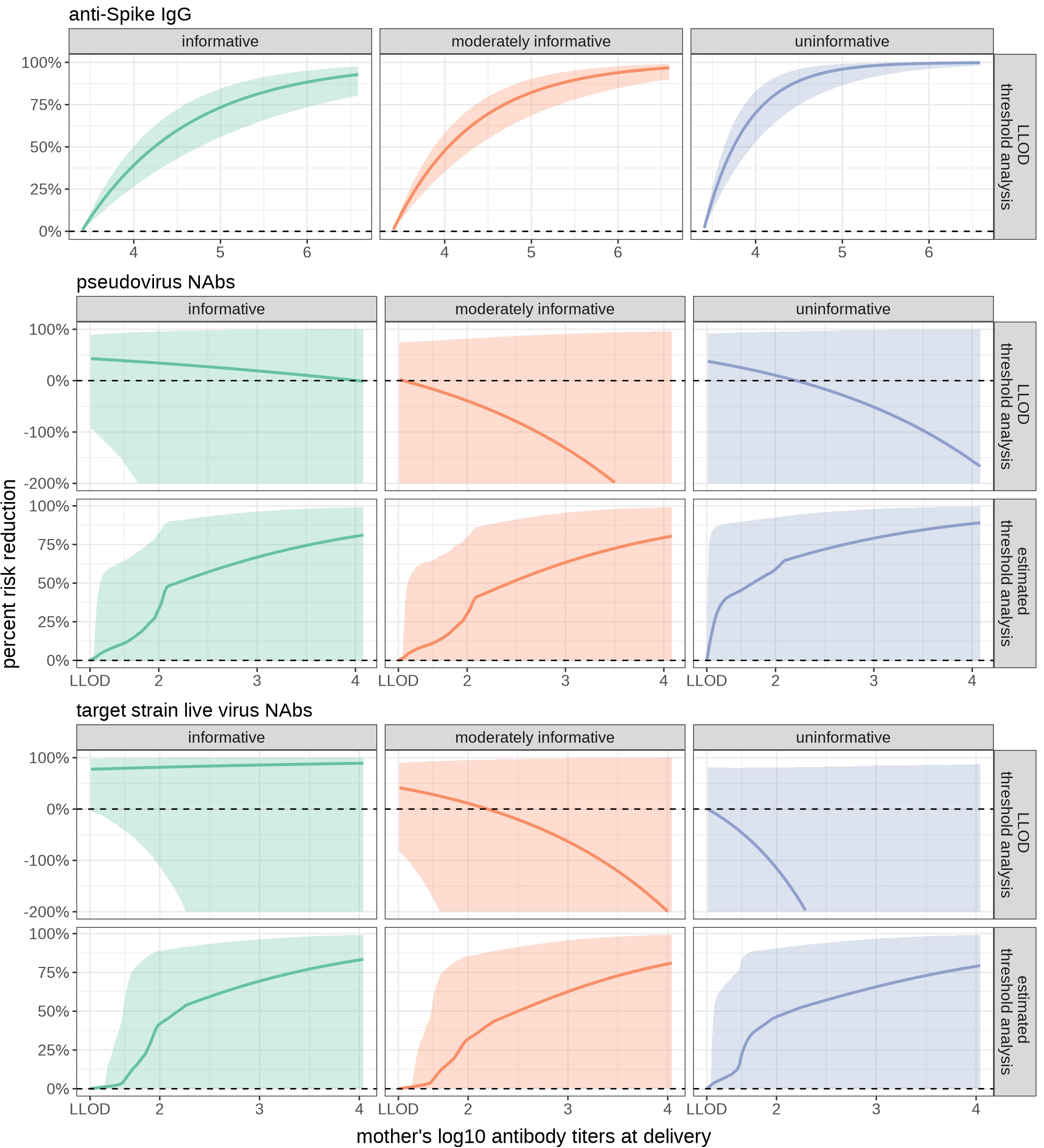}
    \end{center}
    
    \caption{Estimated association between maternal antibodies at delivery and infants' risk of SARS-CoV-2 infection with the Delta variant in the MOMI-Vax study. Posterior means and 95\% posterior credible intervals (CrIs) are shown for the percent reduction in risk of infection in infants compared to the minimum observed maternal antibody titers in the study (top row for each antibody) or compared to the estimated maternal antibody threshold (bottom row for each antibody), adjusted for the covariates included in $\bm{Z}$. All comparisons are among infants at the same site and at the same calendar time. Relative risk reduction is calculated as (1 -- hazard ratio) and is only shown for the observed range of maternal antibodies in the study. Estimates and lower limits of the 95\% CrIs are truncated at -200\% when necessary for readability. The faceting columns refer to the baseline hazard prior used in the model.}
    
    \label{survival:fig:gamma_delta}
\end{figure}

\begin{figure}[H]

    \begin{center}
        \includegraphics[width=1\linewidth]{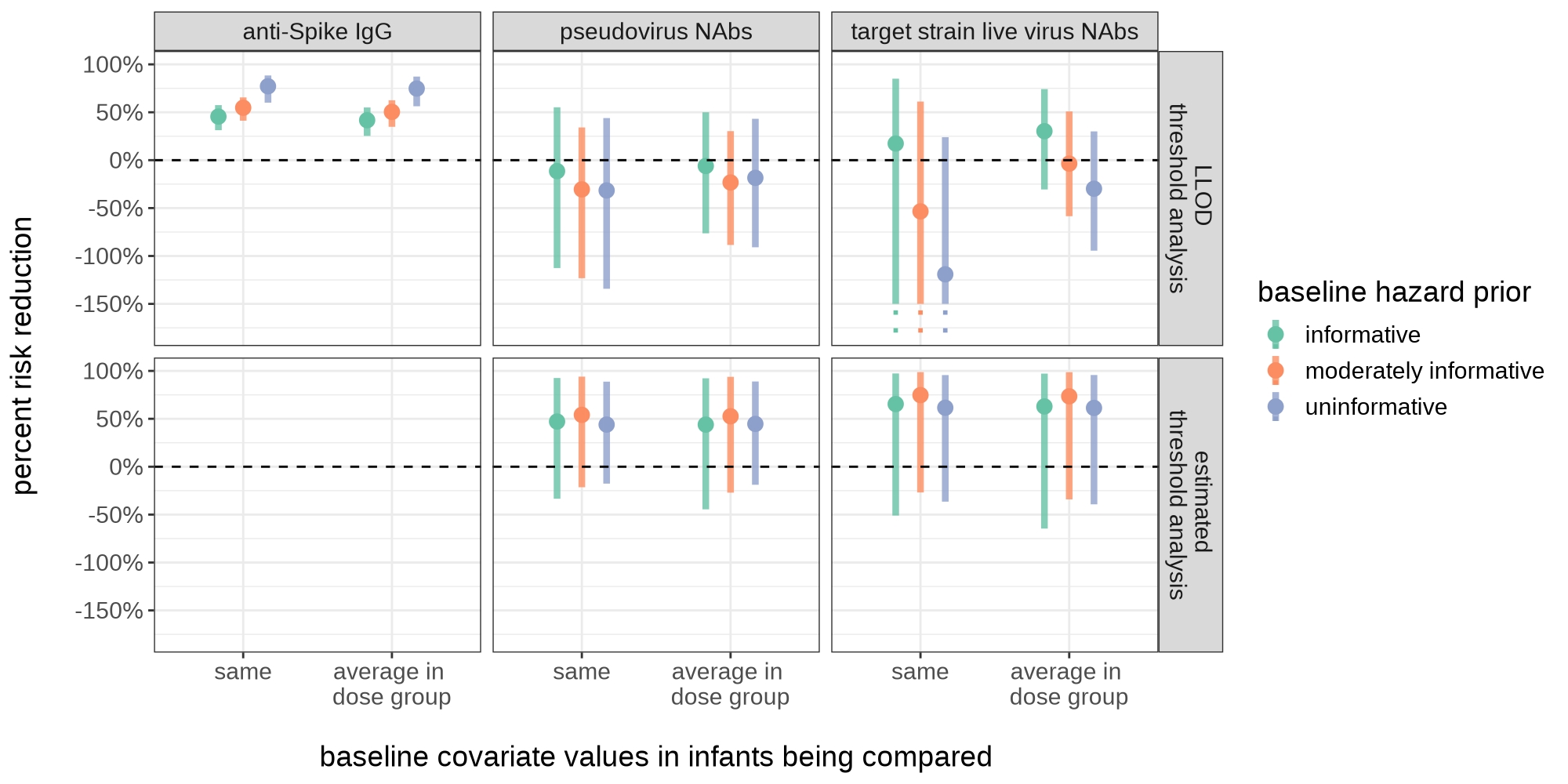}
    \end{center}

    \caption{Estimated relative risk of infection with the SARS-CoV-2 Delta variant among average boosted versus unboosted infants in the MOMI-Vax study. Posterior means and 95\% posterior credible intervals (CrIs) are shown for the percent reduction in risk of infection with the SARS-CoV-2 Delta variant in infants in the MOMI-Vax study, comparing infants whose mothers had the average boosted (3 dose) antibody levels vs. the average nonboosted (2 dose) antibody levels at delivery. Lower limits of the 95\% CrIs are truncated at -150\% when necessary for readability. Relative risk reduction is calculated as (1 -- hazard ratio). For each antibody and baseline hazard prior, results are shown for two comparisons: infants with the same baseline covariates who differ only in their maternal antibody levels (left); and infants whose baseline covariates are set to the average values in their dose group (right), highlighting the differences in risk between the 3 and 2 dose groups beyond calendar time and maternal antibody levels. All comparisons are among infants at the same site and at the same calendar time.}
    \label{survival:fig:gamma_3vs2_delta}
\end{figure}

\afterpage{
\clearpage
\begin{landscape}
\begin{Table}[H]
    \vspace{-0.5in}
    \caption{Results for the association between baseline covariates and the risk of SARS-CoV-2 infection from the MOMI-Vax data analysis.} 
    \begin{center}
        \includegraphics[width=0.7\linewidth]{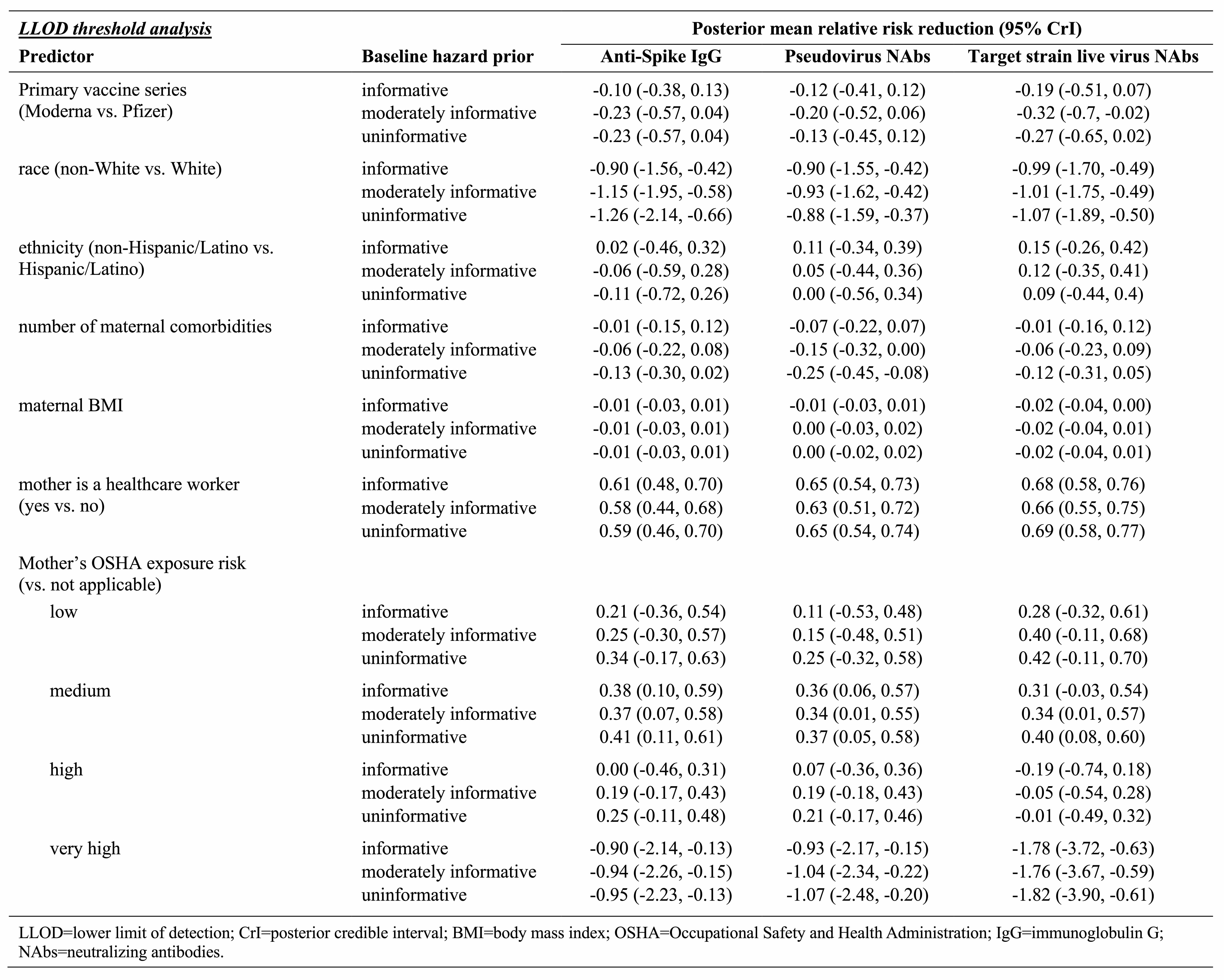}
    \end{center}

    Results are shown for two analyses: one using the lower limit of detection (LLOD) as the threshold below which the model assumes no association between maternal antibodies and the risk of infection, and a second in which this threshold is estimated. Relative risk reduction (RRR) is calculated as (1 -- hazard ratio). Positive RRR means the predictor is associated with a reduced risk of infection, while negative RRR means the predictor is associated with an increased risk of infection.
    \vspace{-0.5in}
    \label{survival:table:momivax_results_beta_h}
\end{Table}

\renewcommand{\theTable}{S\arabic{Table} (continued)}
\begin{Table}[H]\ContinuedFloat
    \vspace{-0.5in}
    \caption[]{Results for the association between baseline covariates and the risk of SARS-CoV-2 infection from the MOMI-Vax data analysis.} 

    \begin{center}
        \includegraphics[width=0.7\linewidth]{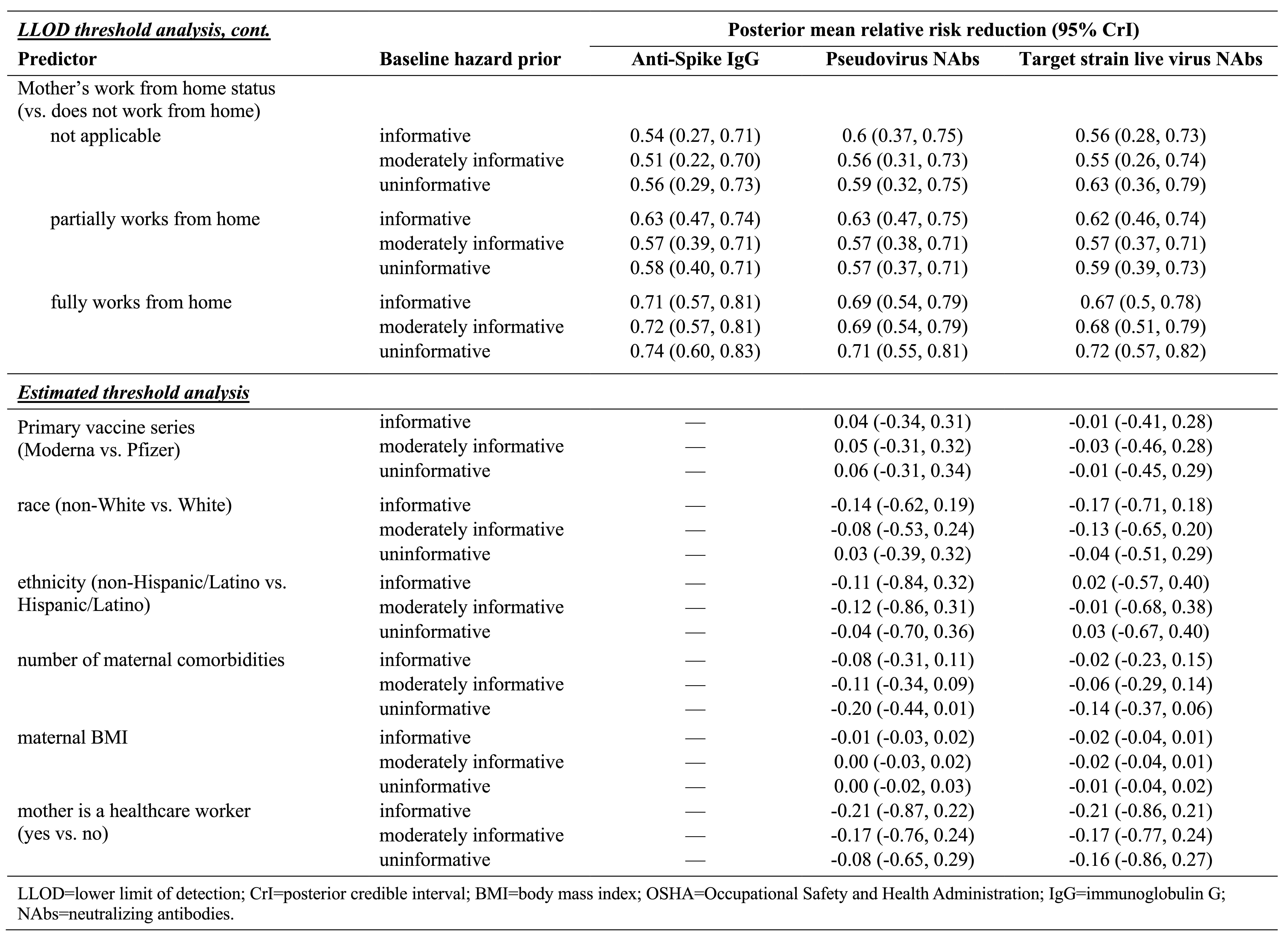}
    \end{center}

    Results are shown for two analyses: one using the lower limit of detection (LLOD) as the threshold below which the model assumes no association between maternal antibodies and the risk of infection, and a second in which this threshold is estimated. Relative risk reduction (RRR) is calculated as (1 -- hazard ratio). Positive RRR means the predictor is associated with a reduced risk of infection, while negative RRR means the predictor is associated with an increased risk of infection.
    
\end{Table}

\begin{Table}[H]\ContinuedFloat
    
    \caption[]{Results for the association between baseline covariates and the risk of SARS-CoV-2 infection from the MOMI-Vax data analysis.} 

    \begin{center}
        \includegraphics[width=0.7\linewidth]{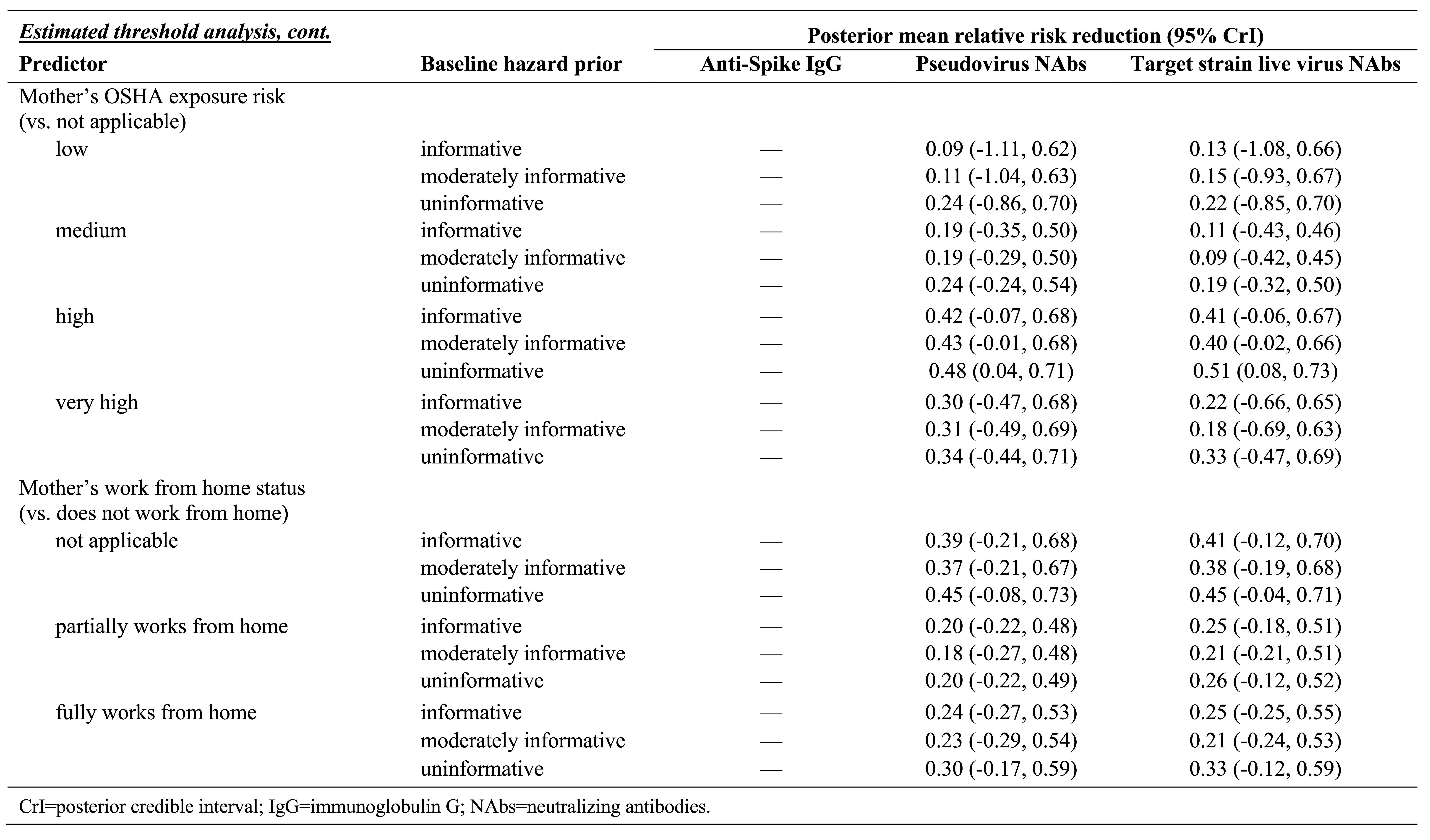}
    \end{center}

    Results are shown for two analyses: one using the lower limit of detection (LLOD) as the threshold below which the model assumes no association between maternal antibodies and the risk of infection, and a second in which this threshold is estimated. Relative risk reduction (RRR) is calculated as (1 -- hazard ratio). Positive RRR means the predictor is associated with a reduced risk of infection, while negative RRR means the predictor is associated with an increased risk of infection.

\end{Table}

\begin{figure}[H]

    \begin{center}
        \includegraphics[width=0.9\linewidth]{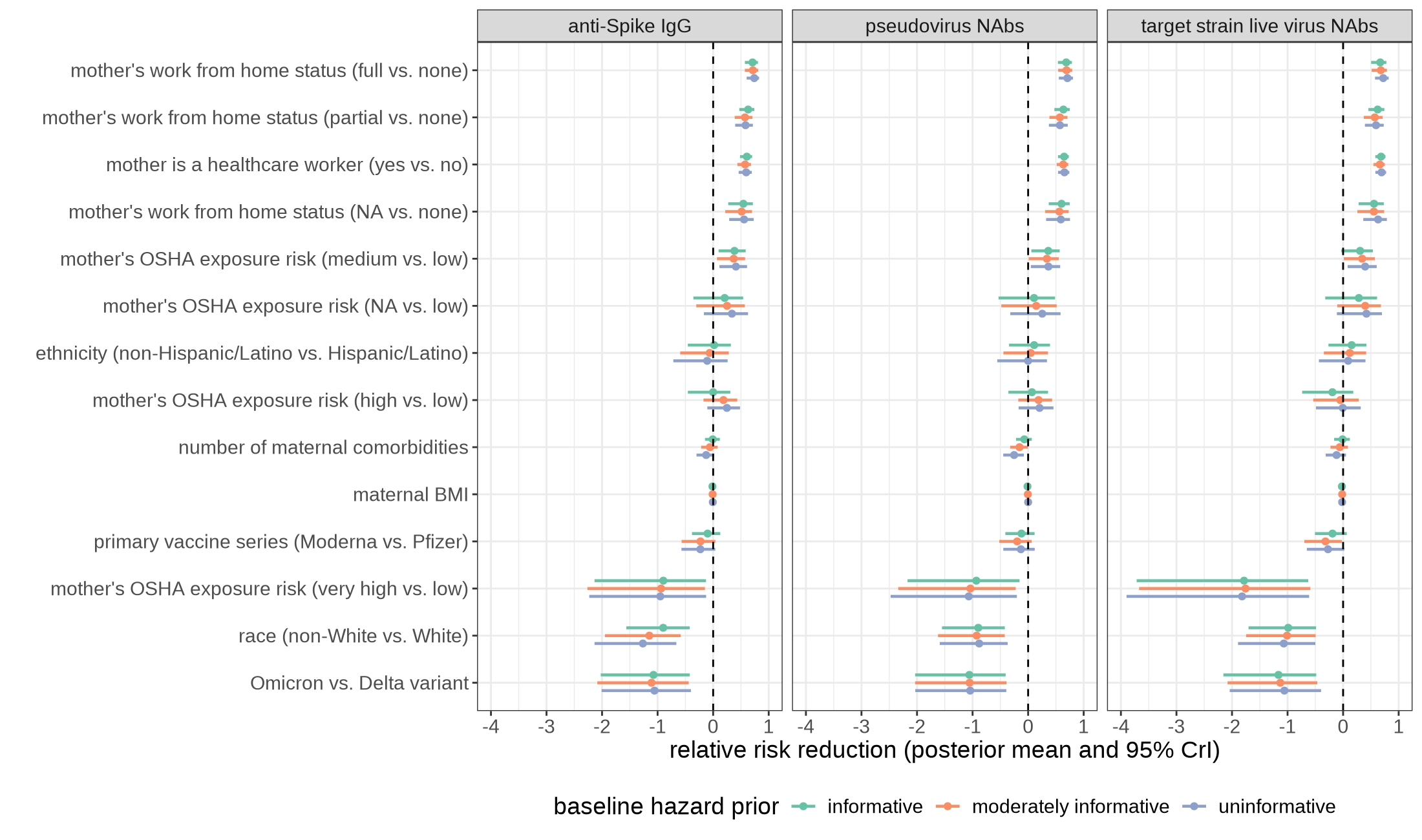}
    \end{center}
    
    \caption{Results for the association between baseline covariates and the risk of SARS-CoV-2 infection from the MOMI-Vax data analysis. Results are shown for two analyses: one using the lower limit of detection (LLOD) as the threshold below which the model assumes no association between maternal antibodies and the risk of infection, and a second in which this threshold is estimated. Relative risk reduction (RRR) is calculated as (1 -- hazard ratio). Positive RRR means the predictor is associated with a reduced risk of infection, while negative RRR means the predictor is associated with an increased risk of infection.}
    
    \label{survival:fig:beta_h}
\end{figure}

\renewcommand{\thefigure}{S\arabic{figure} (continued)}
\begin{figure}[H]\ContinuedFloat

    \begin{center}
        \includegraphics[width=0.9\linewidth]{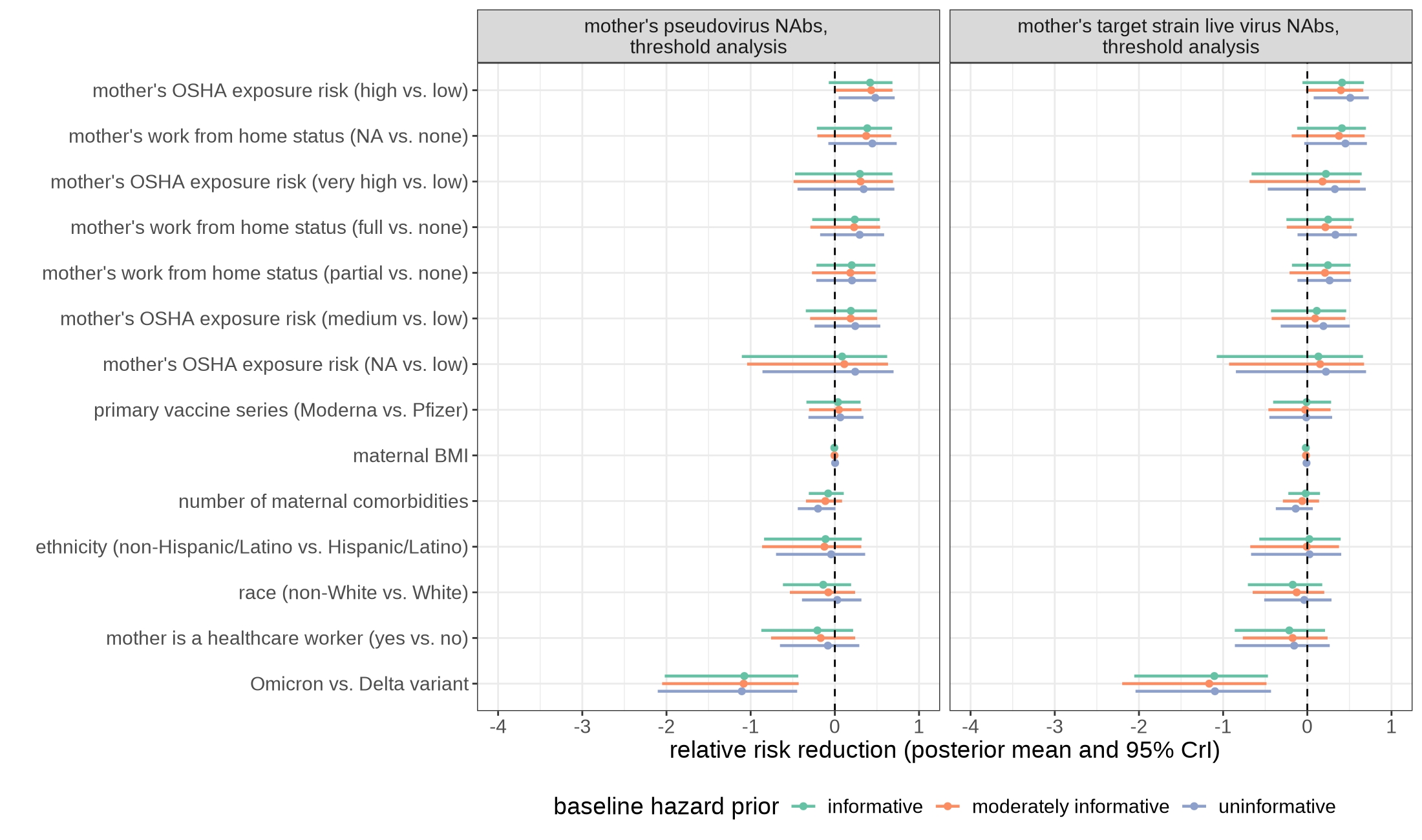}
    \end{center}
    
    \caption{Results for the association between baseline covariates and the risk of SARS-CoV-2 infection from the MOMI-Vax data analysis. Results are shown for two analyses: one using the lower limit of detection (LLOD) as the threshold below which the model assumes no association between maternal antibodies and the risk of infection, and a second in which this threshold is estimated. Relative risk reduction (RRR) is calculated as (1 -- hazard ratio). Positive RRR means the predictor is associated with a reduced risk of infection, while negative RRR means the predictor is associated with an increased risk of infection.}
\end{figure}

\end{landscape}
\clearpage
}

\renewcommand{\thefigure}{S\arabic{figure}}
\begin{figure}[H]

    \begin{center}
        \includegraphics[width=1\linewidth]{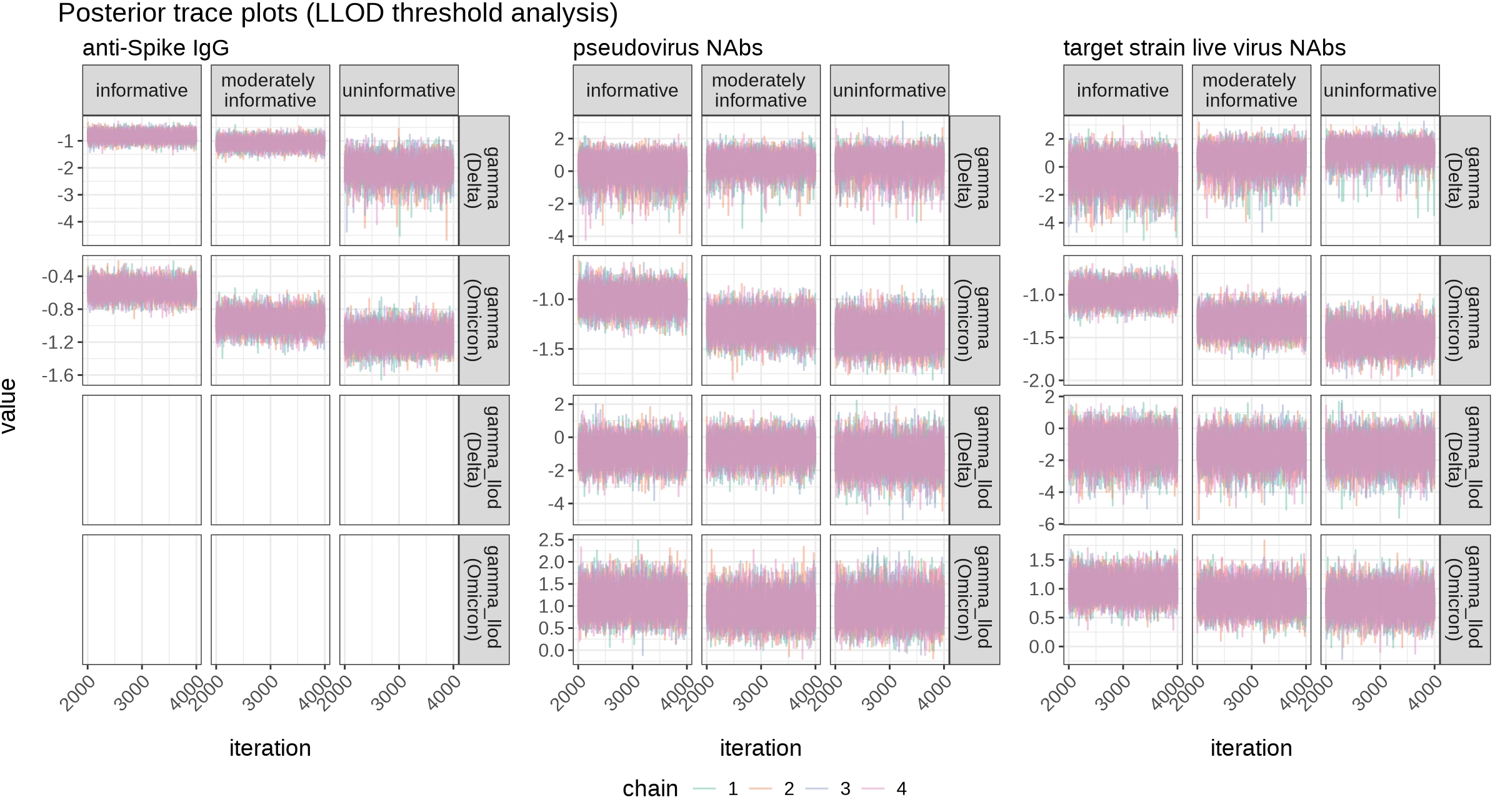} 
    
        \includegraphics[width=1\linewidth]{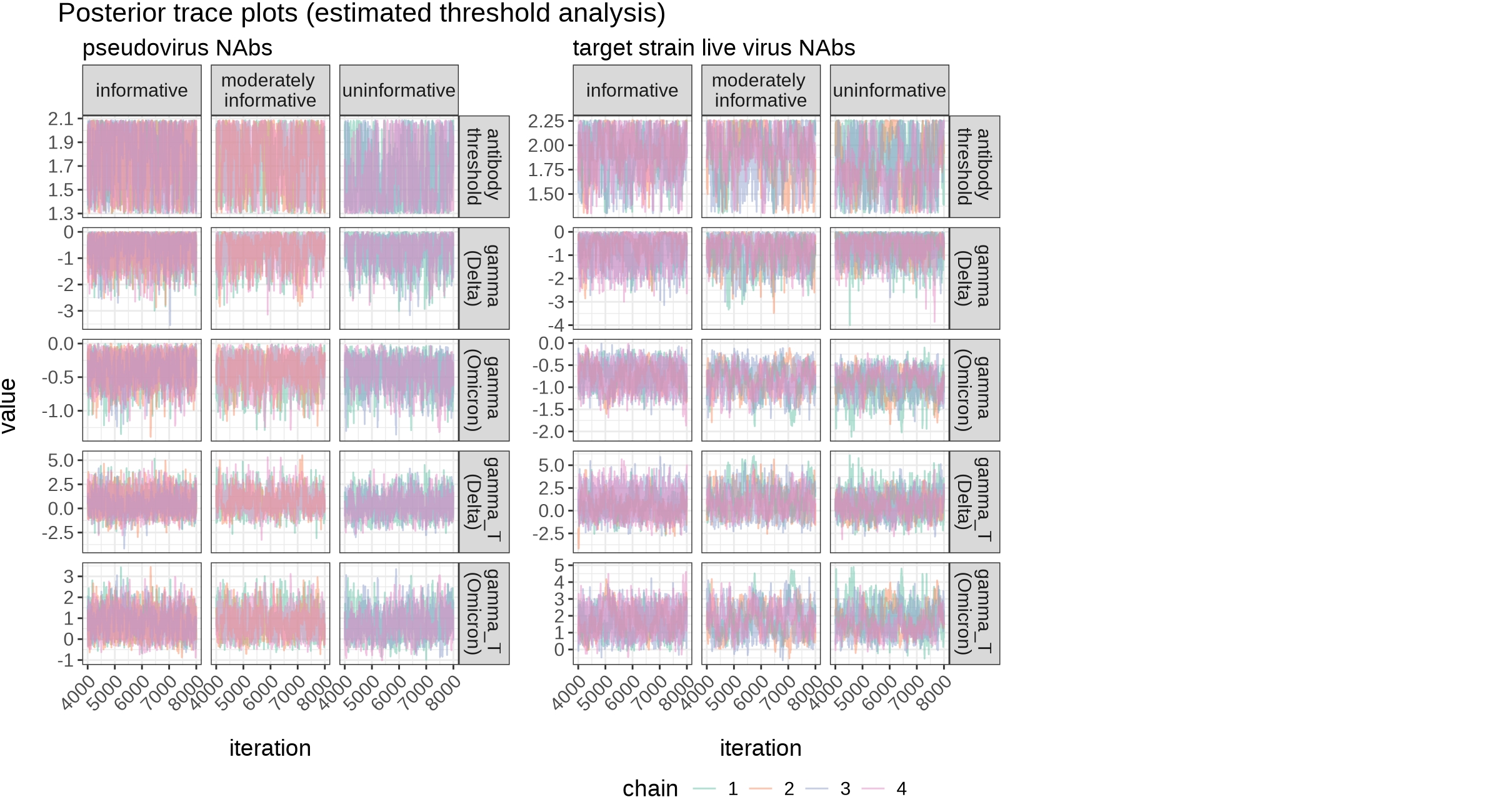} 
    \end{center} 
    
    \caption{Trace plots of the Markov Chain Monte Carlo draws from the posterior for the MOMI-Vax analysis in Stan. For reasons of space, this figure shows trace plots only for the primary parameters of interest: $\gamma^\dvar$, $\gamma^\ovar$, $\gamma_\llod^\dvar$, $\gamma_\llod^\ovar$, $\gamma_T^\dvar$, $\gamma_T^\ovar$, and $X_T$.}
    
    \label{survival:fig:traceplots}
\end{figure}

\begin{figure}[H]
    
    \includegraphics[width=0.9\linewidth]{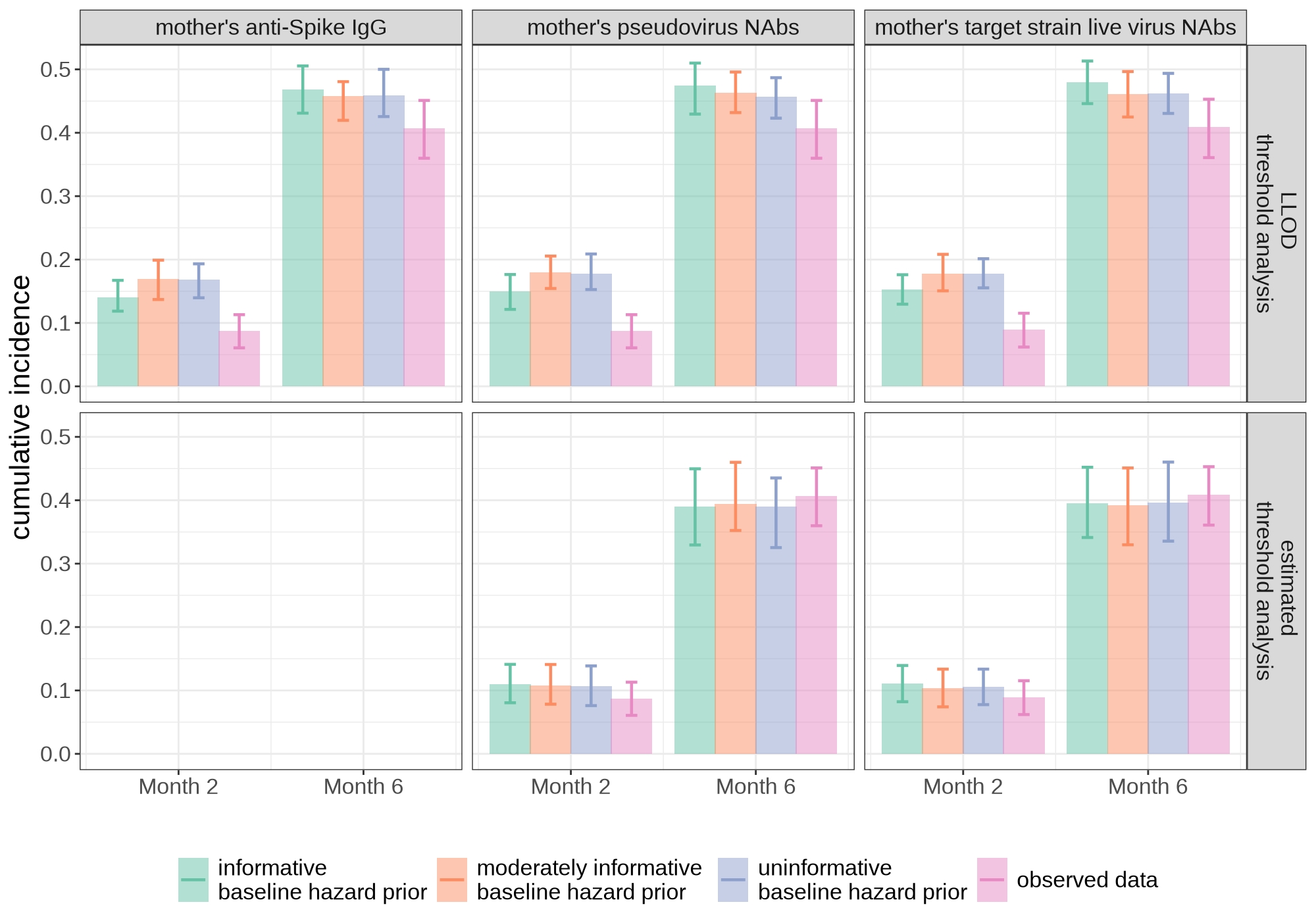}
    
    \caption{Posterior mean cumulative incidence of SARS-CoV-2 infection at Months 2 and 6 from the calendar-time survival model versus observed cumulative incidence in the MOMI-Vax study.}
    
    \label{survival:fig:cuminc}
\end{figure}

\newpage
\begin{code}
    \caption{Stan code used to fit calendar-time survival model to the MOMI-Vax data.}
\end{code}

\footnotesize

\begin{verbatim}

// Code for fitting the calendar time survival model
// This code is for the MOMI-Vax analysis, 
// which includes two variants of interest (Delta "d" and Omicron "o")
// and only baseline covariates in the trajectory and hazard functions

functions {
  real calc_cumhaz(int start_t, int end_t, vector log_bl_hazards, matrix variant_props_data,
                   real alpha_o, real gamma_d, real gamma_o, 
                   real gamma_d_llod, real gamma_o_llod, real X_i, 
                   int ind_llod_i, row_vector Z_i, vector beta_h) {

    int n_days = end_t - start_t + 1;

    vector[n_days] hazard;
    vector[n_days] cumhaz_terms;
    
    for(i in 1:n_days) {
      real prop_hazard_d = variant_props_data[i,1] * exp(gamma_d*X_i*(1-ind_llod_i) + 
                          gamma_d_llod*(1-ind_llod_i));
      real prop_hazard_o = variant_props_data[i,2] * exp(alpha_o + 
                           gamma_o*X_i*(1-ind_llod_i) + gamma_o_llod*(1-ind_llod_i));
      real prop_hazard = (prop_hazard_d + prop_hazard_o) * exp(Z_i*beta_h);
      hazard[i] = exp(log_bl_hazards[i]) * prop_hazard;

      if(i==1) {
        cumhaz_terms[i] = hazard[i];
      }else {
        if(i==n_days) {
          cumhaz_terms[i] = hazard[i];
        }else {
          cumhaz_terms[i] = 2*hazard[i];
        }
      }
    }

    real cumhaz = sum(cumhaz_terms)/2;

    return cumhaz;
  }
}

data {
  ////// DATA //////
  
  // sample size
  int<lower=0> N; 
  
  // predictor variable (tx group, or antibody level at birth)
  vector[N] X;
  
  // baseline covariates for hazard function
  int<lower=0> ncol_Z;
  matrix[N, ncol_Z] Z;
  
  // states (numbered!!!)
  vector[N] states;
  
  // enroll dates (calendar time - relative to 1st day in pw_basis!!!)
  vector[N] enroll_dates;
  
  // birth dates (calendar time - relative to 1st day in pw_basis!!!)
  vector[N] birth_dates;
  
  // visit dates (calendar time - relative to 1st day in pw_basis!!!)
  vector[N] m2_dates;
  vector[N] m6_dates;
  vector[N] max_fu_dates;
  
  // survival times (calendar time - relative to 1st day in pw_basis!!!)
  vector[N] survival_dates_lower;
  vector[N] survival_dates_upper; 
  
  // censoring indicators
  vector[N] ind_right_censored;
  vector[N] ind_int_censored;
  
  // biomarker limit of detection
  real biomarker_llod;
  
  ////// PRIORS //////
  
  // baseline hazard function parameters
  int<lower=0> n_states; 
  int<lower=0> n_pw_coefs;
  
  // vector of indices of the "missing" pw constant coefficient 
  // corresponding to the reference interval
  array[n_states] int<lower=0> reference_coef_num; 
  int<lower=0> n_study_days;
  matrix[n_study_days, n_pw_coefs+1] pw_basis;
  
  // BL hazard during the reference interval
  vector[n_states] log_h0_ref_mean;
  vector[n_states] log_h0_ref_sd;
  
  // pw constant coefficients for BL hazard
  // array of vectors giving the priors for the pw constant coefs for each state
  array[n_states] vector[n_pw_coefs] log_pw_coefs_ratio_mean; 
  array[n_states] vector[n_pw_coefs] log_pw_coefs_ratio_sd;
  
  // relative infectiousness of Omicron variant vs. Delta variant
  real alpha_o_mean;
  real alpha_o_sd;
  
  // data on variant proportions of Delta & Omicron on each study day
  array[n_states] matrix[n_study_days, 2] variant_props_data;
  
  // effect of trajectory function on hazard function (gamma = (gamma_delta, gamma_omicron))
  vector[2] gamma_mean;  
  vector[2] gamma_sd;
  
  vector[2] gamma_llod_mean;  
  vector[2] gamma_llod_sd;
  vector[2] gamma_llod_fixed; // specify values of gamma_llod if we're not estimating them
  
  // effect of baseline covariates on hazard function
  int<lower=0> n_beta_h;
  vector[n_beta_h] beta_h_mean;
  vector[n_beta_h] beta_h_sd;
  
  ////// OPTIONS //////
  
  int<lower=0> estimate_gamma_llod; // 1 = estimate gamma_llod, 0 = do not estimate
  int<lower=0> generate_quantities; // 1 = generate predicted cuminc at M2 & M6; 
                                    // 0 = don't generate
}

parameters {
  vector[n_states] log_h0_ref_raw;
  array[n_states] vector[n_pw_coefs] log_pw_coefs;
  
  real alpha_o_raw;
  vector<upper=0>[2] gamma;
  vector[2] gamma_llod_raw;
  vector[n_beta_h] beta_h_raw;
}

transformed parameters {
  vector[n_states] log_h0_ref = (log_h0_ref_raw .* log_h0_ref_sd) + log_h0_ref_mean;
  
  real alpha_o = (alpha_o_raw*alpha_o_sd) + alpha_o_mean;
  
  real gamma_d = gamma[1]; // delta
  real gamma_o = gamma[2]; // omicron
  
  // only estimate gamma_llod if we specify estimate_gamma_llod
  vector[2] gamma_llod = estimate_gamma_llod ? gamma_llod_raw : gamma_llod_fixed; 
  real gamma_d_llod = gamma_llod[1];
  real gamma_o_llod = gamma_llod[2];
  
  vector[n_beta_h] beta_h = (beta_h_raw .* beta_h_sd) + beta_h_mean;
}

model {
  ////// PRIORS //////
 log_h0_ref_raw ~ std_normal();
  for(i in 1:n_states) {
    log_pw_coefs[i] ~ normal(log_pw_coefs_ratio_mean[i], log_pw_coefs_ratio_sd[i]);
  }

  alpha_o_raw ~ std_normal();
  gamma ~ normal(gamma_mean, gamma_sd);
  beta_h_raw ~ std_normal();
  
  if(estimate_gamma_llod==1){
    gamma_llod_raw ~ std_normal();
  }
  
  // CALCULATE ALL BL HAZARDS //////
  array[n_states] vector[n_pw_coefs+1] log_pw_coefs_with_reference;
  for(i in 1:n_states) {
    log_pw_coefs_with_reference[i][1:(reference_coef_num[i]-1)] = log_h0_ref[i] + 
                                  log_pw_coefs[i][1:(reference_coef_num[i]-1)];
    log_pw_coefs_with_reference[i][reference_coef_num[i]] = log_h0_ref[i];
    log_pw_coefs_with_reference[i][(reference_coef_num[i]+1):(n_pw_coefs+1)] = log_h0_ref[i] + 
                                  log_pw_coefs[i][reference_coef_num[i]:n_pw_coefs];
  }
  
  array[n_states] vector[n_study_days] log_bl_hazards;
  for(i in 1:n_states) {
    log_bl_hazards[i] = pw_basis * log_pw_coefs_with_reference[i];
  }
  
  for(i in 1:N) {
    
    real X_i = X[i];
    row_vector[ncol_Z] Z_i = Z[i,];
    
    int ind_llod_i = 0; // indicator of X_i being below LLOD (0=NOT below LLOD, 1=below LLOD)
    if(X_i <= biomarker_llod) {
      ind_llod_i = 1; 
    }

    int enroll_date_i = to_int(enroll_dates[i]);
    int birth_date_i = to_int(birth_dates[i]);
    int survival_date_lower_i = to_int(survival_dates_lower[i]);
    int survival_date_upper_i = to_int(survival_dates_upper[i]);
    
    real ind_right_censored_i = ind_right_censored[i];
    real ind_int_censored_i = ind_int_censored[i];
    
    // number of days on study (not since birth!!!!)
    int n_days_lower_i = survival_date_lower_i - enroll_date_i + 1; 
    int n_days_upper_i = survival_date_upper_i - enroll_date_i + 1;
    
    // number of days between birth and enroll date
    int days_to_enroll_i = enroll_date_i - birth_date_i; 

    vector[n_days_upper_i] log_bl_hazards_i;
    matrix[n_days_upper_i, 2] variant_props_data_i;
    for(j in 1:n_states){
      if(states[i]==j) {
        log_bl_hazards_i = log_bl_hazards[j][(enroll_date_i+1):(survival_date_upper_i+1)];
        variant_props_data_i = variant_props_data[j][(enroll_date_i+1):(survival_date_upper_i+1),];
      }
    }
    
    if(ind_right_censored_i==0) {
      if(ind_int_censored_i==0) {
        // log-likelihood of infection at time S_i for unensored subjects
        real prop_hazard_d = variant_props_data_i[n_days_upper_i,1] * 
                             exp(gamma_d*X_i*(1-ind_llod_i) + gamma_d_llod*(1-ind_llod_i));
        real prop_hazard_o = variant_props_data_i[n_days_upper_i,2] * exp(alpha_o + 
                             gamma_o*X_i*(1-ind_llod_i) + gamma_o_llod*(1-ind_llod_i));
        target += log_bl_hazards_i[n_days_upper_i] + (Z_i*beta_h) + 
                             log(prop_hazard_d + prop_hazard_o);

      }else {
        // log-likelihood of infection between times L_i and U_i for interval-censored subjects
        real cumhaz_lower_i = calc_cumhaz(enroll_date_i, survival_date_lower_i,
                                          log_bl_hazards_i[1:n_days_lower_i], 
                                          variant_props_data_i[1:n_days_lower_i], 
                                          alpha_o, gamma_d, gamma_o, gamma_d_llod, gamma_o_llod,
                                          X_i, ind_llod_i, Z_i, beta_h);
        real cumhaz_upper_i = calc_cumhaz(enroll_date_i, survival_date_upper_i, 
                                          log_bl_hazards_i, variant_props_data_i, 
                                          alpha_o, gamma_d, gamma_o, gamma_d_llod, gamma_o_llod, 
                                          X_i, ind_llod_i, Z_i, beta_h);
                                          
        // log_diff_exp is a more numerically stable version of 
        // log(exp(-cumhaz_lower)-exp(-cumhaz_upper))                                  
        target += log_diff_exp((-1)*cumhaz_lower_i, (-1)*cumhaz_upper_i); 
      }
    }

    if(ind_int_censored_i==0) {
      // log-likelihood of survival until time S_i for right-censored OR uncensored subjects
      // (but not interval censored!)
      real cumhaz_i = calc_cumhaz(enroll_date_i, survival_date_lower_i, log_bl_hazards_i, 
                                  variant_props_data_i, alpha_o, gamma_d, gamma_o, 
                                  gamma_d_llod, gamma_o_llod, X_i, ind_llod_i, Z_i, beta_h);
      target += (-1)*cumhaz_i;
    }
    
  }
}

generated quantities {
  vector[N] cumhaz_m2;
  vector[N] cumhaz_m6;
  vector[N] delta_m2;
  vector[N] delta_m6;

  // every variable declared inside these brackets are local variables
  if(generate_quantities==1){
    // CALCULATE ALL BL HAZARDS //////
    array[n_states] vector[n_pw_coefs+1] log_pw_coefs_with_reference;
    for(i in 1:n_states) {
      log_pw_coefs_with_reference[i][1:(reference_coef_num[i]-1)] = log_h0_ref[i] + 
                                     log_pw_coefs[i][1:(reference_coef_num[i]-1)];
      log_pw_coefs_with_reference[i][reference_coef_num[i]] = log_h0_ref[i];
      log_pw_coefs_with_reference[i][(reference_coef_num[i]+1):(n_pw_coefs+1)] = log_h0_ref[i] + 
                                     log_pw_coefs[i][reference_coef_num[i]:n_pw_coefs];
    }

    array[n_states] vector[n_study_days] log_bl_hazards;
    for(i in 1:n_states) {
      log_bl_hazards[i] = pw_basis * log_pw_coefs_with_reference[i];
    }

    for(i in 1:N){
      real X_i = X[i];
      row_vector[ncol_Z] Z_i = Z[i,];

      int ind_llod_i = 0; // indicator of X_i being below LLOD (0=NOT below LLOD, 1=below LLOD)
        if(X_i <= biomarker_llod) {
          ind_llod_i = 1;
        }

      int enroll_date_i = to_int(enroll_dates[i]);
      int birth_date_i = to_int(birth_dates[i]);

      int m2_date_i = to_int(m2_dates[i]);
      int n_days_m2_i = m2_date_i - enroll_date_i + 1;

      int m6_date_i = to_int(m6_dates[i]);
      int n_days_m6_i = m6_date_i - enroll_date_i + 1;

      int max_fu_date_i = to_int(max_fu_dates[i]);
      int n_days_max_i = max_fu_date_i - enroll_date_i + 1;

      vector[n_days_max_i] log_bl_hazards_i;
      matrix[n_days_max_i, 2] variant_props_data_i;

      for(j in 1:n_states){
        if(states[i]==j) {
          log_bl_hazards_i = log_bl_hazards[j][(enroll_date_i+1):(max_fu_date_i+1)];
          variant_props_data_i = variant_props_data[j][(enroll_date_i+1):(max_fu_date_i+1),];
        }
      }

      real u_i = uniform_rng(0, 1);

      if(m2_date_i!=999){
        cumhaz_m2[i] = calc_cumhaz(enroll_date_i, m2_date_i, log_bl_hazards_i[1:n_days_m2_i],
                                     variant_props_data_i[1:n_days_m2_i], alpha_o, 
                                     gamma_d, gamma_o,gamma_d_llod, gamma_o_llod, 
                                     X_i, ind_llod_i, Z_i, beta_h);

        real survivor_m2_i = exp(-cumhaz_m2[i]);
        if(u_i > survivor_m2_i){
          delta_m2[i] = 1;
        }else{
          delta_m2[i] = 0;
        }
      }else{
        cumhaz_m2[i] = 999;
        delta_m2[i] = 999;
      }

      if(m6_date_i!=999){
        cumhaz_m6[i] = calc_cumhaz(enroll_date_i, m6_date_i, log_bl_hazards_i[1:n_days_m6_i],
                                     variant_props_data_i[1:n_days_m6_i], alpha_o, 
                                     gamma_d, gamma_o, gamma_d_llod, gamma_o_llod, 
                                     X_i, ind_llod_i, Z_i, beta_h);

        real survivor_m6_i = exp(-cumhaz_m6[i]);
        if(u_i > survivor_m6_i){
          delta_m6[i] = 1;
        }else{
          delta_m6[i] = 0;
        }
      }else{
        cumhaz_m6[i] = 999;
        delta_m6[i] = 999;
      }
    }
  }
}



\end{verbatim}

\end{document}